\documentclass[english]{revtex4}
\usepackage[T1]{fontenc}
\usepackage[latin9]{inputenc}
\usepackage{bm}
\usepackage{amsmath}
\usepackage{graphicx}

\makeatletter

\newcommand{\lyxdot}{.}

\@ifundefined{textcolor}{}
{%
 \definecolor{BLACK}{gray}{0}
 \definecolor{WHITE}{gray}{1}
 \definecolor{RED}{rgb}{1,0,0}
 \definecolor{GREEN}{rgb}{0,1,0}
 \definecolor{BLUE}{rgb}{0,0,1}
 \definecolor{CYAN}{cmyk}{1,0,0,0}
 \definecolor{MAGENTA}{cmyk}{0,1,0,0}
 \definecolor{YELLOW}{cmyk}{0,0,1,0}
}

\makeatother

\usepackage{babel}
\begin{document}

\title{Mode Coupling Theory for Nonequilibrium Glassy Dynamics of Thermal
Self-Propelled Particles}

\author{Mengkai Feng,  Zhonghuai Hou}

\thanks{Corresponding Author: hzhlj@ustc.edu.cn}

\affiliation{Department of Chemical Physics \& Hefei National Laboratory for Physical
Sciences at Microscales, iChEM, University of Science and Technology
of China, Hefei, Anhui 230026, China}

\date{\today}
\begin{abstract}
We present a promising mode coupling theory study for the relaxation
and glassy dynamics of a system of strongly interacting self-propelled
particles, wherein the self-propulsion force is described by Ornstein-Uhlenbeck
colored noise and thermal noises are included. Our starting point
is an effective Smoluchowski equation governing the distribution function
of particle's positions, from which we derive a memory function equation
for the time dependence of density fluctuations in nonequilibrium
steady states. With the basic assumption of absence of macroscopic
currents and standard mode coupling approximation, we can obtain expressions
for the irreducible memory function and other relevant dynamic terms,
wherein the nonequilibrium character of the active system is manifested
through an averaged diffusion coefficient $\bar{D}$ and a nontrivial
structural function $S_{2}\left(q\right)$ with $q$ the magnitude
of wave vector $\mathbf{q}$. $\bar{D}$ and $S_{2}\left(q\right)$
enter the frequency term and the vortex term for the memory function,
thus influence both the short time and the long time dynamics of the
system. With these equations obtained, we study the glassy dynamics
of this thermal self-propelled particles system by investigating the
Debye-Waller factor $f_{q}$ and relaxation time $\tau_{\alpha}$
as functions of the persistence time $\tau_{p}$ of self-propulsion,
the single particle effective temperature $T_{\text{eff}}$ as well
as the number density $\rho$. Consequently, we find the critical
density $\rho_{c}$ for given $\tau_{p}$ shifts to larger values
with increasing magnitude of propulsion force or effective temperature,
in good accordance with previous reported simulation works. In addition,
the theory facilitates us to study the critical effective temperature
$T_{\text{eff}}^{c}$ for fixed $\rho$ as well as its dependence
on $\tau_{p}$. We find that $T_{\text{eff}}^{c}$ increases with
$\tau_{p}$ and in the limit $\tau_{p}\to0$, it approaches the value
for a simple passive Brownian system as expected. Our theory also
well recovers the results for passive systems and can be easily extended
to more complex systems such as active-passive mixtures. 
\end{abstract}
\maketitle

\section{Introduction}

The collective behaviors of systems containing active (self-propelled)
particles have gained extensive attention in recent years due to its
great importance both from a fundamental physics perspective and for
understanding many biological systems \cite{2012_PhysRep_Vicsek_CollectiveMotion,2013_RMP_Marchetti_HydroSoftActi,2016_RMP_Bechinger_ActiParti_in_complex_enviro}.
A wealth of new nonequilibrium phenomena have been reported, such
as active swarming, large scale vortex formation\cite{2010_Nature_Schaller_SoftFilament,2012_Nature_Suminol_HardFilament},
phase separation\cite{2012_PNAS_Linek_PhaseSepa,2012_PRL_Fily_AthermalPhaseSepa,2013_PRL_Buttinoni_ClustPhasSepa,2014_NatComm_Cates_Chi4Theory,2014_PRL_Das_AP_Mix,2016_RMP_Bechinger_ActiParti_in_complex_enviro},
etc, both experimentally and theoretically. Recently, a new trend
in this field has been the glassy dynamics and the glass transition
in dense assemblies of self-propelled particles and their comparisons
to their corresponding phenomena in equilibrium systems\cite{2013_NP_Berthier_NonEquiGlasTrans,2013_NC_NiRan_PushGlasTrans,2013_NatPhys_Kurchan_NonEquiGlasTrans,2014_NatComm_Mandal_ReGT,2014_PRL_Berthier_GTActiveHardDisk,2017_PRE_Brader}.
Experiments demonstrated that active fluids may show dynamic features
such as jamming and dynamic arrest that are very similar to those
observed in glassy materials. For instance, migrating cells exhibited
glassy dynamics, such as dynamic heterogeneity as the cell density
increases\cite{2011_PNAS_Angelini_Exp_Cells}, amorphous solidification
process was found in collective motion of a cellular monolayer\cite{2005_PRL_Krakoviack_porous},
and glassy behaviors could even be found for ant aggregates in large
scales\cite{2002_Science_Whitesides_Self-assemble}, to list just
a few. 

Computer simulations also demonstrated that nonequilibrium glass transition
or dynamic arrest behavior does occur in dense suspensions of self-propelled
particles. So far, mainly two types of self-propulsion particle systems
have been studied. One is the Rotation diffusional Active Brownian
(RAB) particles system, where each particle is subjected to a self-propulsion
force with constant amplitude $v_{0}$ but a randomly changing direction
evolving $via$ rotational diffusion with diffusion coefficient $D_{r}$.
Ni \textit{et al}. \cite{2011_JPCM_Hagen_ActBrown,2013_NC_NiRan_PushGlasTrans,2014_SoftMatt_NiRan_CrystHardSpheGlas}studied
the glassy behavior of this RAB system of hard-sphere particles, finding
that the critical density for glass transition shifts to larger density
as the active force increases, thus pushing the glass transition point
to the limit of random packing. The other one is the so-called Active
Ornstein-Uhlenbeck(AOU) particles system, wherein the self-propulsion
force is realized by a colored noise described by the OU process.
In contrast to the RAB model, thermal noise is ignored in the AOU
system, such that the system can never reach the equilibrium state
determined by the canonical distribution. For this athermal system,
an effective temperature $T_{\text{eff}}$ can be introduced to quantify
the strength of the self-propulsion force, and a persistence time
$\tau_{p}$ controls the duration of the persistent self-propelled
motion. Berthier and co-workers\cite{2014_PRL_Berthier_GTActiveHardDisk,2014_PRE_Berthier_ClustKMCSelfPropHardSph,2016_arXiv_Flenner_NonEquiGlasTrans}
had performed detailed studies about the structural and glassy dynamics
of this AOU model in two and three dimensions. Similarly, the glass
transition shifts to larger densities compared to the equilibrium
one when the magnitude or persistence time of Cthe self-propulsion
force increases. Besides the studies of self-propelled particles,
using molecular dynamics simulations of a model glass former, Mandal
\textit{et al}.\cite{2014_NatComm_Mandal_ReGT} showed that the incorporation
of activity or self-propulsion can induce cage breaking and fluidization,
resulting in a disappearance of the glassy phase beyond a critical
force. And related to the glassy dynamics, it was shown that particle
activity can shift the freezing density to larger values\cite{2012_PRL_Bialke_CrysDenseSuspension}
and particularly, hydrodynamic interactions can further enhance this
effect\cite{2015_SoftMatt_LiShuxian}. 

Besides experimental and simulation studies mentioned above, on the
theoretical side, important progresses have also been achieved in
recent years\cite{2015_our_Arxiv}. Starting from a generalized Langevin
equation with colored non-thermal noise, Berthier and Kurchan\cite{2013_NatPhys_Kurchan_NonEquiGlasTrans}
predicted that dynamic arrest can occur in systems that are far from
equilibrium, showing that non-equilibrium glass transition moves to
lower temperature with increasing activity and to higher temperature
with increasing dissipation in spin glasses. Farage and Brader attempted
to extend the mode coupling theory to the limiting case of the RAB
model\cite{2014_arXiv_Farage_MCTActiveGlass}, wherein activity leads
to a higher effective particle diffusivity. They then started from
the effective Smoluchowski equation governing the many-particle distribution
function and obtained a memory function equation for the equilibrium
intermediate scattering function, showing that self-propulsion could
shift the glass transition to larger density\cite{2015_PRE_Farage_EffectIntera}.
In a recent work, Nandi proposed a phenomenological extension of random
first order transition theory to study glass transition of the RAB
model, showing that more active systems are stronger glass formers
\cite{2016_arXiv_Nandi_Activity}. Very recently, Szamel \textit{et
al.}\cite{2014_PRL_Berthier_GTActiveHardDisk,2015_PRE_Szamel_GlasDyna}
presented an elegant theoretical modeling of the structure and glassy
dynamics of the athermal AOU system. In their approach, they first
integrated out the self-propulsion and then used the projection operator
method and a mode-coupling-like approximation to derive an approximate
equation of motion for the collective intermediate scattering functions,
defined upon the nonequilibrium steady state distribution. In particular,
this work highlighted the importance of the steady state correlations
of particle velocities, which played a crucial role to understand
the relaxation dynamics of the system. Nevertheless, extension of
this framework to more general cases with thermal noise included is
not present yet. 

In the present work, we have developed an alternative mode coupling
theory for the nonequilibrium glass dynamics of a general system of
active particles, wherein the self-propulsion force is described by
an OU process, and besides, thermal noise is included. To be specific,
we call this active OU particles system with thermal noise presented
as the AOU-T system, where 'T' apparently stands for thermal noise.
 Our starting point is an effective Smoluchowski equation(SE) obtained
\textit{via} Fox approximation method, which was recently adopted
by Farage \textit{et al}. to study the effective interactions among
RAB particles\cite{2015_PRE_Farage_EffectIntera}. This effective
SE allows us to derive a memory function equation for the nonequilibrium
steady state collective intermediate scattering function $F_{q}\left(t\right)$,
as well as that for the self-intermediate scattering function $F_{q}^{s}\left(t\right)$.
With the basic assumptions of absence of macroscopic currents and
standard mode coupling approximation, we are able to get the expressions
for the irreducible memory functions and other relevant variables.
Particularly, we find that the dynamics is governed by an averaged
diffusion coefficient $\bar{D}$ and a nontrivial steady state structure
function $S_{2}\left(q\right)$, both depending on the effective temperature
$T_{\text{eff}}$, the persistence time $\tau_{p}$ as well as the
number density $\rho$. With $\bar{D}$, $S_{2}\left(q\right)$ and
the nonequilibrium structure factor $S\left(q\right)$ as inputs,
we can calculate $F_{q}\left(t\right)$ and $F_{q}^{s}\left(t\right)$
for different parameter settings and investigate the glass transition
behaviors. We calculate the critical density $\rho_{c}$ for glass
transition as a function of the effective temperature $T_{\text{eff}}$,
the magnitude of propulsion force $v_{0}$ and the persistence time
$\tau_{p}$, by investigating the Debye-Waller factor $f_{q}$ as
well as the relaxation time $\tau_{\alpha}$. Consequently, $\rho_{c}$
shifts to larger values with increasing propulsion force $v_{0}$
or effective temperature $T_{\text{eff}}$, in good accordance with
previous simulation works. The theory also facilitates us to study
the critical temperature $T_{\text{eff}}^{c}$ for fixed number density
$\rho$, as well as its dependence on $\tau_{p}$. 

The remainder of paper is organized as follows. In Section II, we
present descriptions of the AOU-T model and our theory, the latter
being the main result of the present work. Section III includes the
numerical results predicted by our theory and conclusions in Section
IV. 

\section{Model and Theory\label{sec:Model-and-Theory}}

\subsection{AOU-T Model for self-propelled particles}

We consider a system of $N$ interacting, self-propelled particles
in a volume $V$. The particles move in a viscous medium with single
particle friction coefficient $\gamma$ and hydrodynamic interactions
are neglected. As mentioned in the introduction, the self-propulsion
force is given by a colored noise described by OU process and thermal
noises are included. The equations of motion for these AOU-T particles
are thus given by

\begin{align}
\dot{\mathbf{r}}_{i}\left(t\right) & =\gamma^{-1}\left[\mathbf{F}_{i}\left(t\right)+\mathbf{f}_{i}\left(t\right)\right]+\bm{\xi}_{i}\left(t\right)\label{eq:AOU-T-1}
\end{align}
where $\mathbf{r}_{i}$ denotes the position vector of particle $i$,
the force $\mathbf{F}_{i}=-\sum_{j\ne i}\nabla_{i}u\left(r_{ij}\right)$
originates from the interactions where is the pair-potential $u\left(r_{ij}\right)$,
$\mathbf{f}_{i}$ is the self-propulsion force, and $\bm{\xi}_{i}\left(t\right)$
is the thermal noise with zero mean and variance 
\begin{equation}
\left\langle \bm{\xi}_{i}\left(t\right)\bm{\xi}_{j}\left(t'\right)\right\rangle =2D_{t}\bm{1}\delta_{ij}\delta\left(t-t'\right),\label{eq:xi_t}
\end{equation}
where $D_{t}=k_{B}T/\gamma$ with $k_{B}$ the Boltzmann constant
and $T$ the ambient temperature, and $\bm{1}$ denotes the unit tensor.
The equations of motion for the self-propulsion force $\mathbf{f}_{i}$
are given by

\begin{equation}
\dot{\mathbf{f}_{i}}\left(t\right)=-\tau_{p}^{-1}\mathbf{f}_{i}\left(t\right)+\bm{\eta}{}_{i}(t)\label{eq:fi_t}
\end{equation}
where $\tau_{p}$ is the persistence time of self-propulsion and $\bm{\eta}{}_{i}(t)$
is a Gaussian white noise with zero mean and variance $D_{f}$

\begin{equation}
\left\langle \bm{\eta}_{i}\left(t\right)\bm{\eta}_{j}\left(t'\right)\right\rangle =2D_{f}\mathbf{1}\delta_{ij}\delta\left(t-t'\right)\label{eq:etai_t}
\end{equation}
Accordingly, the correlation function of the force $\mathbf{f}_{i}$
reads
\begin{equation}
\left\langle \mathbf{f}_{i}(t)\mathbf{f}_{j}(t')\right\rangle =D_{f}\tau_{p}e^{-\left|t'-t\right|/\tau_{p}}\mathbf{1}\delta_{ij}\label{eq:fi_corr}
\end{equation}

For an isolated particle, the mean square displacement can be obtained
as
\begin{equation}
\left\langle \delta r^{2}\left(t\right)\right\rangle =\frac{6D_{f}\tau_{p}^{2}}{\gamma^{2}}\left[t+\tau_{p}\left(e^{-t/\tau_{p}}-1\right)\right]+6D_{t}t\label{eq:delta_r2t}
\end{equation}
In the short time limit $t\ll\tau_{p}$, the particle's motion is
diffusive with $\left\langle \delta r^{2}\left(t\right)\right\rangle =6D_{t}t$,
which is at variance with the AOU system, wherein the term $6D_{t}t$
is absent and the motion is ballistic $\left\langle \delta r^{2}\left(t\right)\right\rangle =3D_{f}\tau_{p}\gamma^{-2}t^{2}$
for $t\ll\tau_{p}$. For long time, the motion is also diffusive but
with $\left\langle \delta r^{2}\left(t\right)\right\rangle =6\left(D_{t}+D_{f}\tau_{p}^{2}/\gamma^{2}\right)t$,
implying that the long diffusion coefficient is given by 
\begin{equation}
D_{0}=D_{t}+D_{f}\tau_{p}^{2}/\gamma^{2}\label{eq:D_0}
\end{equation}
This allows us to introduce a single-particle effective temperature
\begin{equation}
T_{\text{eff}}=T+k_{B}D_{f}\tau_{p}^{2}/\gamma=\left(\frac{D_{0}}{D_{t}}\right)T.\label{eq:Teff}
\end{equation}
In the limit of vanishing $\tau_{p}\rightarrow0$, $\left\langle \mathbf{f}_{i}(t)\mathbf{f}_{j}(t')\right\rangle \rightarrow2D_{f}\tau_{p}^{2}\bm{1}\delta\left(t-t'\right)\delta_{ij}$
and the system becomes equivalent to a Brownian system at the effective
temperature $T_{\text{eff}}$ with diffusion coefficient $D_{0}$.
For the AOU model where the thermal noise is absent, the effective
temperature is simply given by $T_{\text{{\normalcolor eff}}}=D_{f}\tau_{p}^{2}/\gamma$\cite{2014_PRE_Szamel_EffecTemp}
. 

We note here that the AOU-T model can be mapped onto the RAB model
at a coarse-grained level\cite{2015_PRE_Farage_EffectIntera}. The
equation of motion for the RAB particles is $\dot{{\bf r}}_{i}=v_{0}{\bf p}_{i}+\gamma^{-1}{\bf F}_{i}+\bm{\xi}_{i}$,
where $v_{0}$ denotes the magnitude of the propulsion force and $\mathbf{p}_{i}$
is the unit vector of the direction of particle $i$. $\mathbf{p}_{i}$
changes randomly with time \textit{via} rotational diffusion, $\dot{{\bf p}}_{i}=\bm{\zeta}_{i}\times{\bf p}_{i}$,
where $\bm{\zeta}_{i}$ is also a Gaussian white noise with zero mean
and correlation $\left\langle \bm{\zeta}_{i}\left(t\right)\bm{\zeta}_{j}\left(t'\right)\right\rangle =2D_{r}\bm{1}\delta_{ij}\delta\left(t-t'\right)$
where $D_{r}$ denotes the rotational diffusion coefficient. Recently,
it was shown that $\mathbf{p}_{i}$ can be approximated by a colored
noise with persistence time $\tau_{p}=\left(2D_{r}\right)^{-1}$ if
one average over the angular degree of freedom, i.e., 
\begin{equation}
\left\langle \mathbf{p}_{i}\left(t\right)\mathbf{p}_{j}\left(t'\right)\right\rangle \simeq\frac{1}{3}e^{-2D_{r}\left|t-t'\right|}{\bf 1}\delta_{ij}\label{eq:pi_corr}
\end{equation}
Comparing Eq.(\ref{eq:pi_corr}) with (\ref{eq:fi_corr}), we see
that $v_{0}\mathbf{p}_{i}$ has same correlation property as $\mathbf{f}_{i}$
by the mapping $\tau_{p}\to\left(2D_{r}\right)^{-1}$ and $D_{f}\tau_{p}\to v_{0}^{2}/3$.

In the studies of the AOU model, the authors usually used $T_{\text{eff}}$
as one of the independent parameters together with the persistence
time $\tau_{p}$ and number density $\rho=N/V$. In the simulation
works of the RAB model, however, the authors often used the magnitude
of the propulsion force $v_{0}$ as an independent parameter. Note
that in the simulation work performed by Ran Ni \textit{et al}.\cite{2013_NC_NiRan_PushGlasTrans},
they have adopted Stokes-Einstein relation to set the rotational diffusion
coefficient $D_{r}=3D_{t}/\sigma^{2}$ where $\sigma$ is the particle
diameter. In the dimensionless version by setting $D_{t}=1$, $\gamma=1$
and $\sigma=1$, this means that $D_{r}$ is fixed and the persistence
time is $\tau_{p}=\left(2D_{r}\right)^{-1}=1/6$. In more general
cases, the coupling of $D_{t}$ and $D_{r}$ may not hold and one
can thus set $\tau_{p}$ as a free independent parameter. In the present
work, we will set $v_{0}=\sqrt{3D_{f}\tau_{p}}$ or $T_{\text{eff}}$,
$\tau_{p}$, and $\rho$ as independent parameters if not otherwise
specified. 

For simplicity, we consider a one-component pure-repulsive Lenard-Jones(LJ)
system of self-propelled particles. The pair potential is given by
\begin{equation}
u\left(r\right)=\begin{cases}
4\varepsilon\left[\left(\frac{\sigma}{r}\right)^{12}-\left(\frac{\sigma}{r}\right)^{6}\right]+\varepsilon & r\leq2^{1/6}\sigma\\
0 & r>2^{1/6}\sigma
\end{cases}\label{eq:u_r}
\end{equation}
where $\varepsilon$ is the strength of the potential. Here we set
$\epsilon=1k_{B}T$, where $k_{B}T=D_{t}\gamma$ is the unit of energy.
Moreover, $\sigma\gamma/k_{B}T$ is the unit of time. The number density
$\rho$ is set to be large enough such that the phase separation dynamics
is not relevant and we mainly focus on the glassy dynamics. Simulations
are performed in a cubic box with $L=10$ and periodic boundary conditions.

\subsection{Effective Smoluchowski Equation}

The AOU-T model described in Eq.(\ref{eq:AOU-T-1}) is non-Markovian
due to the colored noise term $\mathbf{f}_{i}$. Consequently, it
is not possible to derive an exact Fokker-Planck equation (FPE) for
the time evolution of the probability distribution function $\Psi\left(\mathbf{r}^{N},t\right)$,
which gives the probability that the system is at a specific configuration
$\mathbf{r}^{N}=\left(\mathbf{r}_{1},\mathbf{r}_{2},\dots,\mathbf{r}_{N}\right)$
at time $t$. Nevertheless, one may obtain an approximate FPE for
such a colored noise system by applying the method introduced by Fox\cite{1986_PRA_Fox},
where a perturbative expansion in powers of correlation time is partially
resumed using functional calculus. The resulting FPE thus defines
implicitly a Markovian process, and is shown to be rather accurate
for short correlation time of the colored noise. Very recently, T.
Farage and co-workers had applied such a method to the RAB model and
obtained an effective FPE for $\Psi\left(\mathbf{r}^{N},t\right)$,
and analyzed the effective interaction among active particles in the
low density limit. Since the FPE only involves the distribution in
the configuration space $\mathbf{r}^{N}$, it is also known as Smoluchowski
equation (SE). 

Here we use the same method to obtain the approximate SE of the AOU-T
model. The procedure is similar to that in Ref.\cite{2015_PRE_Farage_EffectIntera}
noting that the AOU-T model can be mapped to the RAB model as discussed
in the last subsection. For self-consistency, the main steps with
necessary illustrations are given in Appendix A. Finally, we can obtain
the effective SE as 

\begin{equation}
\frac{\partial}{\partial t}\Psi\left(\mathbf{r}^{N},t\right)=\hat{\Omega}\Psi\left(\mathbf{r}^{N},t\right)\label{eq:SmoFunc}
\end{equation}
where $\hat{\Omega}$ denotes the effective Smoluchowski operator
given by
\begin{equation}
\hat{\Omega}=\sum_{j=1}^{N}\nabla_{j}\cdot D_{j}(\mathbf{r}^{N})\left[\nabla_{j}-\beta\mathbf{F}_{j}^{{\rm eff}}\left(\mathbf{r}^{N}\right)\right].\label{eq:SmoOpe}
\end{equation}
Herein, $D_{j}\left(\mathbf{r}^{N}\right)$ is a configuration-dependent
instantaneous diffusion coefficient of particle $j$ which is given
by 
\begin{equation}
D_{j}\left(\mathbf{r}^{N}\right)=D_{t}+\frac{D_{f}\tau_{p}^{2}/\gamma^{2}}{1-\tau_{p}\beta D_{t}\nabla_{j}\cdot\mathbf{F}_{j}^{\mbox{ }}}.\label{eq:D_j}
\end{equation}
with $\beta=\left(k_{B}T\right)^{-1}$. $\mathbf{F}_{j}^{\text{eff}}\left(\mathbf{r}^{N}\right)$
defines an instantaneous effective force subject to particle $j$,
which also depends on the configuration, given by

\begin{equation}
\mathbf{F}_{j}^{\mbox{eff}}\left(\mathbf{r}^{N}\right)=\frac{D_{t}}{D_{j}\left(\mathbf{r}^{N}\right)}\left[\mathbf{F}_{j}\left(\mathbf{r}^{N}\right)-\frac{1}{\beta D_{t}}\nabla_{j}D_{j}\left(\mathbf{r}^{N}\right)\right]\label{eq:F_eff}
\end{equation}

Note that for passive particles, one has $D_{f}\tau_{p}=0$ such that
$D_{j}\left(\mathbf{r}^{N}\right)=D_{t}$ and $\mathbf{F}_{j}^{\text{eff}}\left(\mathbf{r}^{N}\right)=\mathbf{F}_{j}\left(\mathbf{r}^{N}\right)$
as expected. In the limit $\tau_{p}\to0$,corresponding to a white
noise $\mathbf{f}_{i}$, we have $D_{j}\left(\mathbf{r}^{N}\right)=D_{t}+D_{f}\tau_{p}^{2}/\gamma^{2}=D_{0}$
and $\mathbf{F}_{j}^{\text{eff}}=\left(D_{t}/D_{0}\right)\mathbf{F}_{j}$.
In this latter case, the effective Smoluchowski operator is given
by\cite{2014_arXiv_Farage_MCTActiveGlass}
\begin{equation}
\hat{\Omega}_{\tau_{p\to0}}=D_{0}\sum_{j=1}^{N}\nabla_{j}\left(\nabla_{j}-\beta\frac{D_{t}}{D_{0}}\mathbf{F}_{j}\right)=D_{0}\sum_{j=1}^{N}\nabla_{j}\left(\nabla_{j}-\beta_{\text{eff}}\mathbf{F}_{j}\right)\label{eq:SO_Tau0}
\end{equation}
where $\beta_{\text{eff}}=\left(k_{B}T_{\text{eff}}\right)^{-1}$,
and the system reduces to $N$ interacting Brownian particles at the
effective temperature $T_{\text{eff}}$. 

We assume that the system will reach a nonequilibrium steady state
(NESS) $P_{s}\left(\mathbf{r}^{N}\right)$ in the long time limit,
which satisfies 
\begin{equation}
\hat{\Omega}P_{s}\left(\mathbf{r}^{N}\right)=-\sum_{i}\nabla_{i}\cdot\mathbf{J}_{i}^{s}=0\label{eq:Omega_Pss}
\end{equation}
where the steady state current $\mathbf{J}_{i}^{s}$ is given by 
\begin{equation}
\mathbf{J}_{i}^{s}=-D_{j}\left(\mathbf{r}^{N}\right)\left[\nabla_{j}-\beta\mathbf{F}_{j}^{{\rm eff}}\left(\mathbf{r}^{N}\right)\right]P_{s}\left(\mathbf{r}^{N}\right)\label{eq:J_i_s}
\end{equation}
For a passive system, $P_{s}\left(\mathbf{r}^{N}\right)$ will be
given by the canonical equilibrium distribution $P_{s}^{\text{eq}}\left(\mathbf{r}^{N}\right)=\exp\left(-\beta U\left(\mathbf{r}^{N}\right)\right)/Z$
where $U\left(\mathbf{r}^{N}\right)=\frac{1}{2}\sum_{j\ne i}u\left(r_{ij}\right)$
is the system potential and $Z$ is the partition function. But for
the active system studied in the present work, the explicit form of
$P_{s}\left(\mathbf{r}^{N}\right)$ is hard to obtain. Nevertheless,
in the case $\tau_{p}\to0$, $P_{s}^{\tau_{p}\to0}\left(\mathbf{r}^{N}\right)\sim\exp\left(-\beta_{\text{eff}}U\left(\mathbf{r}^{N}\right)\right)$
satisfies $\hat{\Omega}_{\tau_{p\to0}}P_{s}^{\tau_{p}\to0}\left(\mathbf{r}^{N}\right)=0$
indicating that the system can be described by an effective equilibrium
distribution at an effective temperature $T_{\text{eff}}$. 

For latter purposes, it is convenient to introduce an adjoint operator
of the Smoluchowski operator as 
\[
\hat{\Omega}^{\dagger}=\sum_{j=1}^{N}\left(\nabla_{j}+\beta\mathbf{F}_{j}^{{\rm eff}}\right)D_{j}\left(\mathbf{r}^{N}\right)\cdot\nabla_{j}
\]
which satisfies $\int d\mathbf{r}^{N}f^{*}\left(\hat{\Omega}g\right)=\int d\mathbf{r}^{N}\left(\hat{\Omega}^{\dagger}f\right)^{*}g$
for any functions $f\left(\mathbf{r}^{N}\right)$ and $g\left(\mathbf{r}^{N}\right)$.
For the collective dynamic behaviors of the system, one can then
define the collective intermediate scattering function as\cite{2014_arXiv_Farage_MCTActiveGlass}

\begin{equation}
F_{q}\left(t\right)=\frac{1}{N}\left\langle \text{\ensuremath{\rho}}_{\mathbf{q}}^{*}\left(e^{\hat{\Omega}^{\dagger}t}\rho_{\mathbf{q}}\right)\right\rangle =\frac{1}{N}\left\langle \text{\ensuremath{\rho}}_{-\mathbf{q}}\left(e^{\hat{\Omega}^{\dagger}t}\rho_{\mathbf{q}}\right)\right\rangle \label{eq:Fq_t}
\end{equation}
where 
\begin{equation}
\rho_{{\bf q}}=\sum_{j=1}^{N}e^{-i{\bf q}\cdot{\bf r}_{j}}\label{eq:Rho_q}
\end{equation}
is the Fourier transform with wave vector $\mathbf{q}$ of the density
variable $\rho\left(\mathbf{r}\right)=\sum_{j=1}^{N}\delta\left(\mathbf{r}-\mathbf{r}_{j}\right)$
and $q=\left|\mathbf{q}\right|$. In particular, one must emphasize
that the brackets $\left\langle \right\rangle $ in Eq.(\ref{eq:Fq_t})
denotes the ensemble average over the NESS distribution $P_{s}\left(\mathbf{r}^{N}\right)$,
rather than over the equilibrium one $P_{s}^{\text{eq}}\left(\mathbf{r}^{N}\right)$.
For $t=0$, $F_{q}\left(t\right)$ is related to the non-equilibrium
static structure factor 
\begin{equation}
F_{q}\left(0\right)=\frac{1}{N}\left\langle \text{\ensuremath{\rho}}_{-\mathbf{q}}\rho_{\mathbf{q}}\right\rangle =S\left(q\right)\label{eq:NE_Sq}
\end{equation}
where again $\left\langle \cdots\right\rangle $ denotes averaging
over the NESS. Nevertheless, for the non-equilibrium system studied
here, $S\left(q\right)$ can not be calculated by analytical methods
like the Ornstein-Zernike (OZ) equations and must be obtained by direct
simulations. 

Note that $F_{q}\left(t\right)$ can also be written as 
\begin{equation}
F_{q}\left(t\right)=\frac{1}{N}\left\langle \text{\ensuremath{\rho}}_{-\mathbf{q}}e^{\hat{\Omega}t}\rho_{\mathbf{q}}\right\rangle \label{eq:Fq_t_Omega}
\end{equation}
wherein the operator $\hat{\Omega}$ acts on all the functions on
its right side including $P_{s}\left(\mathbf{r}^{N}\right)$, while
in Eq.(\ref{eq:Fq_t}) the adjoint operator $\hat{\Omega}^{\dagger}$
only acts on $\rho_{\mathbf{q}}$. We also consider a closely related
function, $F_{q}^{s}\left(t\right)$, called self-intermediate scattering
function
\begin{eqnarray}
F_{q}^{s}\left(t\right) & = & \left\langle \rho_{-q}^{s}\rho_{q}^{s}\left(t\right)\right\rangle =\left\langle e^{-i\mathbf{q}\cdot\left(\mathbf{r}_{s}(t)-\mathbf{r}_{s}(0)\right)}\right\rangle \label{eq:Fsq_t}\\
 & = & \frac{1}{N}\sum_{j=1}^{N}\left\langle e^{-i\mathbf{q}\cdot\left(\mathbf{r}_{j}(t)-\mathbf{r}_{j}(0)\right)}\right\rangle \nonumber \\
 & = & \frac{1}{N}\sum_{j=1}^{N}\left\langle \text{\ensuremath{\rho}}_{-\mathbf{q}}^{j}e^{\hat{\Omega}t}\rho_{\mathbf{q}}^{j}\right\rangle 
\end{eqnarray}
where $\rho_{q}^{s}$ is the Fourier transform of microscopic tagged
particle (tracer) density $\rho_{q}^{s}=e^{-i\mathbf{q}\cdot\mathbf{r}_{s}}$. 

\subsection{Memory Function Equations}

In this subsection, we derive a formal expression for the collective
(and self-) intermediate scattering functions Eqs. (\ref{eq:Fq_t_Omega})
and (\ref{eq:Fsq_t}) in terms of the so-called irreducible memory
function. This can be done most easily in the Laplace domain, and
the details are given in the Appendix B. Consequently, the equation
for the time evolution of $F_{q}\left(t\right)$ is given by
\begin{equation}
\frac{\partial}{\partial t}F_{q}(t)+\omega_{q}F_{q}(t)+\int_{0}^{t}du\tilde{M}^{{\rm irr}}\left(q,t-u\right)\frac{\partial}{\partial u}F_{q}(u)=0\label{eq:GLE_Fq}
\end{equation}
where 
\begin{equation}
\omega_{q}=-\left\langle \rho_{\mathbf{q}}^{*}\left(\hat{\mbox{\ensuremath{\Omega}}}^{\dagger}\rho_{\mathbf{q}}\right)\right\rangle \left\langle \rho_{\mathbf{q}}^{*}\rho_{\mathbf{q}}\right\rangle ^{-1}=\frac{q^{2}\sum_{j}\left\langle D_{j}\left(\mathbf{r}^{N}\right)\right\rangle }{NS(q)}=\frac{q^{2}\bar{D}}{S\left(q\right)}\label{eq:Omega_q}
\end{equation}
is the frequency term where $D_{j}\left(\mathbf{r}^{N}\right)$ is
given by Eq.(\ref{eq:D_j}) and
\begin{equation}
\bar{D}=N^{-1}\sum_{j}\left\langle D_{j}\left(\mathbf{r}^{N}\right)\right\rangle .\label{eq:Alpha}
\end{equation}
denotes an averaged single-particle diffusion coefficient in the NESS. 

The irreducible memory function $\tilde{M}^{{\rm irr}}\left(q,t\right)$
is given by 
\begin{equation}
\tilde{M}^{{\rm irr}}\left(q,t\right)=\frac{\rho\bar{D}}{16\pi^{3}}\int d\mathbf{k}\left[\left(\hat{{\bf q}}\cdot\mathbf{k}\right)C_{2}\left(\mathbf{q};\mathbf{k}\right)+\left(\hat{{\bf q}}\cdot\mathbf{p}\right)C_{2}\left(\mathbf{q};\mathbf{k}\right)\right]^{2}F_{k}\left(t\right)F_{p}\left(t\right).\label{eq:Memory_Irr}
\end{equation}
with $\mathbf{p}=\mathbf{q}-\mathbf{k}$, $p=\left|\mathbf{p}\right|$.
Herein, a pseudo-correlation function $C_{2}\left(\mathbf{q},\mathbf{k}\right)$
is introduced which is defined as 
\begin{equation}
C_{2}\left(\mathbf{q};\mathbf{k}\right)=\rho^{-1}\left[1-\frac{D_{0}}{\bar{D}}\frac{S_{2}\left(p\right)}{S\left(p\right)}S^{-1}\left(k\right)\right]\label{eq:C2_qk}
\end{equation}
where
\begin{equation}
S_{2}(k)=\frac{1}{D_{0}N}\left\langle \sum_{i,j}D_{j}\left(\mathbf{r}^{N}\right)e^{-i\mathbf{k}\cdot\mathbf{r}_{j}+i\mathbf{k}\cdot\mathbf{r}_{i}}\right\rangle \label{eq:S2_k}
\end{equation}
denotes a static structure function involving the coupling of the
instantaneous diffusion coefficient $D_{j}\left(\mathbf{r}^{N}\right)$
and density fluctuation $e^{-i\mathbf{k}\cdot\left(\mathbf{r}_{j}-\mathbf{r}_{i}\right)}$. 

Eqs. (\ref{eq:GLE_Fq}) to (\ref{eq:S2_k}) contribute to the main
theoretical results of the present paper. The equation for $F_{q}\left(t\right)$,
(\ref{eq:GLE_Fq}), has the same form as that for an equilibrium colloid
system\cite{1996_PhysRep_Nagele_charged_suspensions}. However, Eqs.(\ref{eq:Omega_q})
to (\ref{eq:S2_k}) contain important new features that are specific
to the AOU-T system. The frequency $\omega_{q}$ depends on the parameter
$\bar{D}$, which denotes an averaged effective diffusion coefficient
of a particle in the NESS. The irreducible memory function, Eq.(\ref{eq:Memory_Irr}),
has similar form as that for passive colloid systems, except that
a new pseudo-direct correlation function $C_{2}\left(\mathbf{q};\mathbf{k}\right)$
is introduced in replace of the usual direct correlation function
$c(k)=\rho^{-1}\left[1-S^{-1}\left(k\right)\right]$. The definition
of $C_{2}\left(\mathbf{q};\mathbf{k}\right)$ now involves another
function $S_{2}\left(k\right)$, defined by Eq.(\ref{eq:S2_k}), which
resembles the structure factor $S\left(k\right)$ but with $D_{j}\left(\mathbf{r}^{N}\right)$
involved. Since $D_{j}$ is a configuration-dependent function, it
cannot be drawn out of the summation $\sum_{i,j}$ in Eq.(\ref{eq:S2_k}).
Interestingly, for a homogeneous passive system, $D_{j}=D_{t}=D_{0}$,
hence $\omega_{q}=q^{2}D_{t}S^{-1}\left(q\right)$, $S_{2}\left(k\right)=N^{-1}\left\langle \sum_{i,j}e^{-i\mathbf{k}\cdot\mathbf{r}_{j}+i\mathbf{k}\cdot\mathbf{r}_{i}}\right\rangle \equiv S\left(k\right)$
and $C_{2}\left(\mathbf{q};\mathbf{k}\right)$ simply reduces to $c\left(k\right)$.
In this case, Eq. (\ref{eq:Memory_Irr}) becomes 
\begin{equation}
\tilde{M}^{irr}\left(q,t\right)=\frac{\rho D_{t}}{16\pi^{3}}\int d\mathbf{k}\left[\left(\hat{{\bf q}}\cdot\mathbf{k}\right)c\left(k\right)+\left(\hat{{\bf q}}\cdot\mathbf{p}\right)c\left(k\right)\right]^{2}F_{k}\left(t\right)F_{p}\left(t\right)
\end{equation}
which reduces exactly to that of a passive colloid system\cite{1987_PhysA_Cichocki_irreducible,1994_PhysA_Kawasaki_irreducible}.
Note that in the limit $\tau_{p}\to0$, we have $D_{j}=\overline{D}=D_{0}$
such that $S_{2}\left(k\right)=S\left(k\right)$ and $C_{2}\left(\mathbf{q};\mathbf{k}\right)=c\left(k\right)$
also hold. In this case, $\omega_{q}=q^{2}D_{0}S^{-1}\left(q\right)$
and the equations describe the dynamics of an equivalent Brownian
system with effective diffusion coefficient $D_{0}$ as described
above. 

In general cases, $D_{j}$ is dependent on the particle positions,
thus it cannot be drawn out from the summation of $S_{2}\left(k\right)$
in Eq.(\ref{eq:S2_k}). Interestingly, if we approximately replaces
$D_{j}$ by its ensemble average value $\left\langle D_{j}\right\rangle $
in the summation, we can obtain
\begin{align}
S_{2}\left(k\right) & \simeq\frac{1}{D_{0}N}\left\langle \sum_{i,j}\left\langle D_{j}\right\rangle e^{-i\mathbf{k}\cdot\mathbf{r}_{j}+i\mathbf{k}\cdot\mathbf{r}_{i}}\right\rangle \label{eq:S2_k_Sk}\\
 & =\frac{\bar{D}}{D_{0}N}\left\langle \sum_{i,j}e^{-i\mathbf{k}\cdot\mathbf{r}_{j}+i\mathbf{k}\cdot\mathbf{r}_{i}}\right\rangle =\frac{\bar{D}}{D_{0}}S\left(k\right)\nonumber 
\end{align}
in the second equality, we use the fact that $\left\langle D_{j}\right\rangle =\bar{D}$
for a homogenous system. We find then 
\begin{equation}
C_{2}\left(\mathbf{q},\mathbf{k}\right)=\rho^{-1}\left[1-\frac{D_{0}}{\bar{D}}\frac{S_{2}\left(p\right)}{S\left(p\right)}S^{-1}\left(k\right)\right]\simeq c\left(k\right)
\end{equation}
and the memory function Eq.(\ref{eq:Memory_Irr}) reduces to that
for a passive system with effective diffusion coefficient $\bar{D}$.
Note that this approximation holds if the coupling of $D_{j}$ and
density fluctuation $e^{-i{\bf k}\cdot\left({\bf r}_{j}-\mathbf{r}_{l}\right)}$
is weak or the fluctuation of $D_{j}$ is very small. In the latter
sections of the present paper, we will show by simulations that $S_{2}\left(k\right)\simeq\left(\bar{D}/D_{0}\right)S\left(k\right)$
is a very good approximation when $k$ is large, whereas for small
$k$ $S_{2}\left(k\right)D_{0}/\bar{D}S\left(k\right)$ does show
apparent structures.

In the next section, we will adopt our above theoretical results to
study the glassy behaviors of the one-component LJ active system described
by the AOU-T model. In the dimensionless unit, $\gamma=D_{t}=k_{B}T=1$
and we choose $D_{f}$, $\tau_{p}$ together with the number density
$\rho$ as adjustable parameters. As already discussed in the model
description part, now the effective temperature is given by $T_{\text{eff}}=1+D_{f}\tau_{p}^{2}$,
while the amplitude of active force is quantified by $v_{0}=\sqrt{3D_{f}\tau_{p}}$.
To compare our results with those simulation works of Ni and others,
we will choose $v_{0}$ and $\tau_{p}$ as independent variables together
with $\rho$. Nevertheless, we will also study the behavior of the
system by choosing $T_{\text{eff}}$ and $\tau_{p}$ as independent
free parameters since it has been a regular choice in recent studies\cite{2015_PRE_Szamel_GlasDyna,2016_arXiv_Flenner_NonEquiGlasTrans}.
To begin, we will run the system until it reaches the steady state
from which we can get the parameter $\bar{D}$ and the function $S_{2}\left(k\right)$,
with which the memory function Eq. (\ref{eq:GLE_Fq}) can be numerically
calculated. We can then investigate the time dependence of $F_{q}\left(t\right)$
to address the glass transition issue. 

For the self-intermediate scattering function $F_{q}^{s}\left(t\right)$,
the memory function equation reads(see the Appendix \ref{appGLE})
\begin{equation}
\frac{\partial}{\partial t}F_{q}^{s}(t)+\omega_{q}^{s}F_{q}^{s}(t)+\int_{0}^{t}duM_{s}^{{\rm irr}}\left(q,t-u\right)\frac{\partial}{\partial u}F_{q}^{s}(u)=0\label{eq:GLE_Fsqt}
\end{equation}
where $\omega_{q}^{s}=q^{2}\bar{D}$ and 
\begin{equation}
\tilde{M}_{s}^{{\rm irr}}\left(q;t\right)=\frac{\rho\bar{D}}{\left(2\pi\right)^{3}}\int d^{3}\mathbf{k}\left[\left(\mathbf{k}\cdot\hat{\mathbf{q}}\right)c(k)+\left(\mathbf{p}\cdot\hat{\mathbf{q}}\right)\frac{1}{\rho}\left(1-\frac{D_{0}S_{2}\left(k\right)}{\bar{D}S(k)}\right)\right]^{2}F_{k}(t)F_{p}^{s}(t)\label{eq:M_irr_s}
\end{equation}

If $S_{2}\left(k\right)\simeq\left(\bar{D}/D_{0}\right)S\left(k\right)$,
the second term in the bracket can be neglected, and the equation
reduces to the equilibrium version. 

It would be instructive here to compare our theoretical results with
those in the literature. As mentioned in the introduction, Farage
and Brader\cite{2014_arXiv_Farage_MCTActiveGlass} had tried to develop
a MCT for the RAB model in the limit $\tau_{p}\to0$. In this circumstance,
the effective Smoluchowski operator is actually given by Eq.(\ref{eq:SO_Tau0}).
Starting from this effective operator, they obtained a memory function
for the collective scattering function $F_{q}^{\text{eq}}\left(t\right)$,
but defined for the equilibrium distribution. In our work, the effective
Smoluchowski operator is now extended to finite (small) $\tau_{p}$,
and importantly, the scattering function is now defined for the NESS
which is more relevant for the active system as pointed out by Szamel\cite{2015_PRE_Szamel_GlasDyna}.
The extension to finite $\tau_{p}$ and using a nonequilibrium function
makes it feasible to compare with simulation results. Surely, for
a nonequilibrium MCT theory, some static functions such as $\bar{D}$,
$S_{2}\left(k\right)$ in the present work must be obtained from simulations,
which is currently not avoidable. 

On the other hand, Szamel \textit{et al}.\cite{2016_PRE_Szamel_AthermalActi}
had made important progress in the theoretical modeling of active
particle systems very recently. In particular, they mainly focused
on athermal system, the AOU model, which is applicable for large colloidal
systems wherein thermal noise might be ignored compared to the self-propulsion.
Their treatment followed a quite different way as in the present work,
where they performed a projection onto the local steady state defined
by the self-propulsion force $\mathbf{f}_{i}$. With the assumption
of vanishment of system currents in the local steady state and mode
coupling approximation, they were able to obtain an effective Smoluchowski
operator, which is time dependent, and the memory function for the
nonequilibrium scattering function $F_{q}\left(t\right)$. Importantly,
their theory involved a function $\omega_{||}\left(q\right)$ which
highlights the role of the velocity correlations. In particular, this
theory reproduced a nontrivial non-monotonic dependence of the relaxation
time $\tau_{\alpha}$ with $\tau_{p}$ if $T_{\text{eff}}$ is fixed
which was observed in their simulations for a standard LJ system,
although the theory apparently overestimated $\tau_{\alpha}$ in the
$\tau_{p}\to0$ limit. In our present work, we have considered the
AOU-T model where thermal noise is taken into account. We have not
tried to extend Szamel's method to this thermal situation, which might
be hard to realize, rather we have adopted a different scheme. Given
that the Fox's method is applicable, the effective Smoluchowski operator
given by Eq.(\ref{eq:SmoOpe}) provides the starting point for the
derivation. This approach actually involves a type of coarse-gaining
over time, wherein the effects of colored noise has been replaced
by an effective white one but with configuration-dependent correlation
functions. As shown in our theory above, the dynamics is then mainly
determined by the effective diffusion coefficient $\bar{D}$ and a
static structure function $S_{2}\left(k\right)$ wherein both involves
the instantaneous diffusion coefficient $D_{j}\left(\mathbf{r}^{N}\right)$.
Interestingly, although our method are quite different with that of
Szamel, we note that $\omega_{||}\left(q\right)\tau_{p}$ in their
work plays the same role as $\bar{D}$ in ours. We also note that
in a recent paper, Marconi \textit{et al}.\cite{2015_ScitiRep_Maggi_Multidimen_Stationary}
had studied the velocity correlations in the AOU-model, finding an
expression very similar to $D_{j}\left(\mathbf{r}^{N}\right)$ under
mean-field approximation. 

\section{Numerical Results and Discussions\label{sec:Result-and-Discussions}}

\subsection{Static Properties }

\begin{figure}
\begin{centering}
\par\end{centering}
\begin{centering}
\includegraphics[width=0.8\columnwidth]{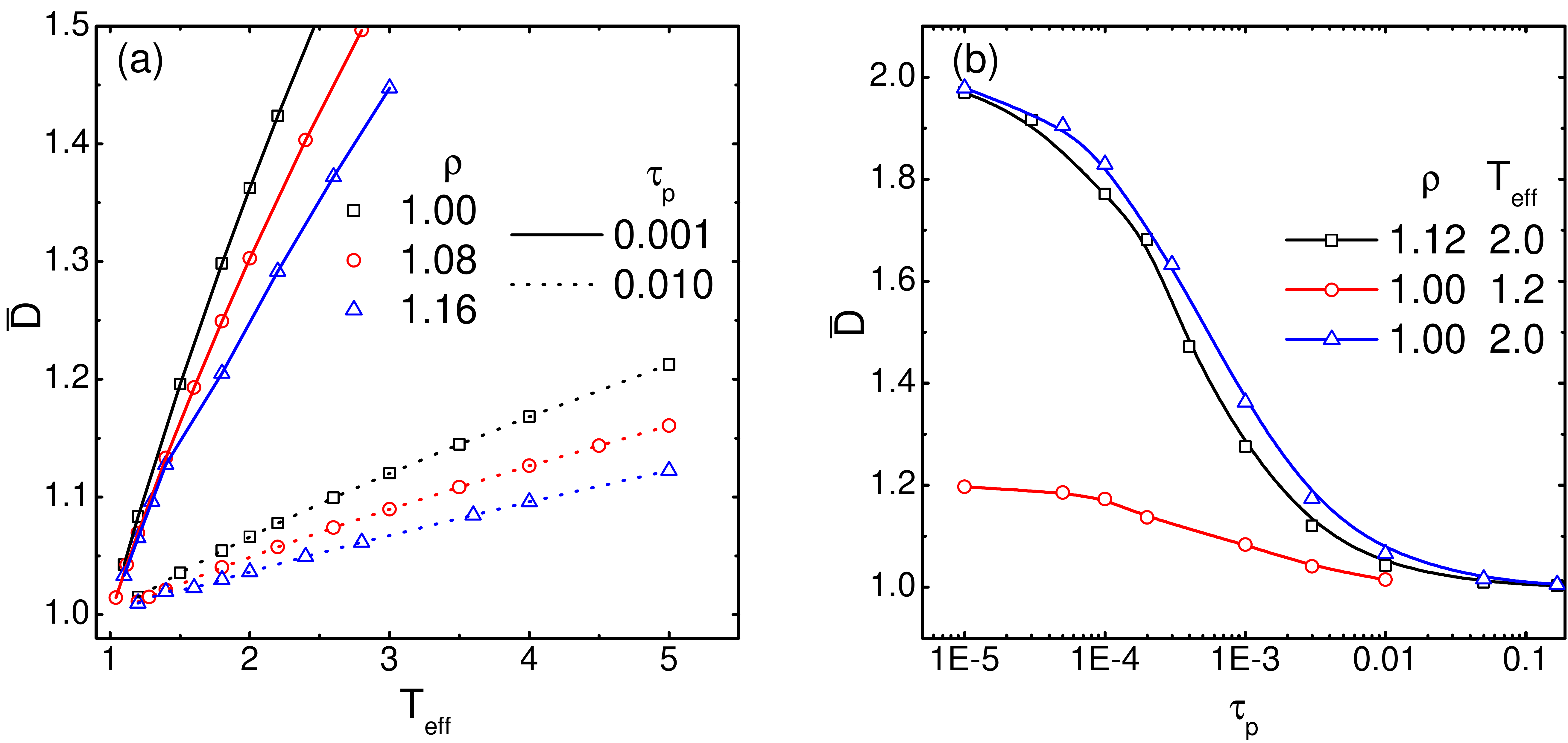}
\par\end{centering}

\caption{The dependence of averaged diffusion coefficient $\bar{D}$ on the
density $\rho$, effective temperature $T_{\text{eff}}$ and persistent
time $\tau_{p}$. (a) Parameter $\bar{D}$ displays a monotonic increasing
with $T_{{\rm eff}}$, at all $\tau_{p}=$0.001(solid line), 0.01(dot
line), and for $\rho=$1.00(squares), 1.08(circles), 1.16(trigonals).
(b) The variances of $\bar{D}$ with $\tau_{p}$ , for different $\rho$
and $T_{{\rm eff}}$.}
\label{Fig_Dbar}
\end{figure}

As discussed above, to solve $F_{q}\left(t\right)$, we must obtain
the parameter $\bar{D}$ and the pseudo-structure factor $S_{2}\left(k\right)$
in the NESS $via$ direct numerical simulations. In Fig.\ref{Fig_Dbar}(a),
the dependence of $\bar{D}$ on the effective temperature $T_{\text{eff}}$
is presented, for different fixed values of $\tau_{p}$ and $\rho$.
As can be seen, $\bar{D}$ increases monotonically with $T_{\text{eff}}$,
which is reasonable since $\bar{D}$ denotes a kind of averaged diffusion
coefficient that should be larger for a higher temperature. If $T_{\text{eff}}$
is fixed, $\bar{D}$ decreases with increasing $\tau_{p}$ and the
variation of $\bar{D}$ with $T_{\text{eff}}$ becomes less sharp,
i.e., $\left(\partial\bar{D}/\partial T_{\text{eff}}\right)$ decreases.
Such qualitative behaviors are robust with the change of number density
$\rho$, despite that the value of $\bar{D}$ decreases slightly with
increasing $\rho$ for given values of $\tau_{p}$ and $T_{\text{eff}}$.
In Fig.\ref{Fig_Dbar}(b), we have also plotted $\bar{D}$ as a function
of persistent time $\tau_{p}$ for different values of $T_{{\rm eff}}$
and $\rho$. In this case, we can see that $\bar{D}$ decreases with
$\tau_{p}$, tending to approach $D_{0}=k_{B}T_{{\rm eff}}/\gamma$
when $\tau_{p}\rightarrow0$ and close to $D_{t}$ at a large $\tau_{p}$
value. Besides, at the lower density $\bar{D}$ has a slightly larger
value as shown in (a). 

\begin{figure}
\begin{centering}
\includegraphics[width=0.6\columnwidth]{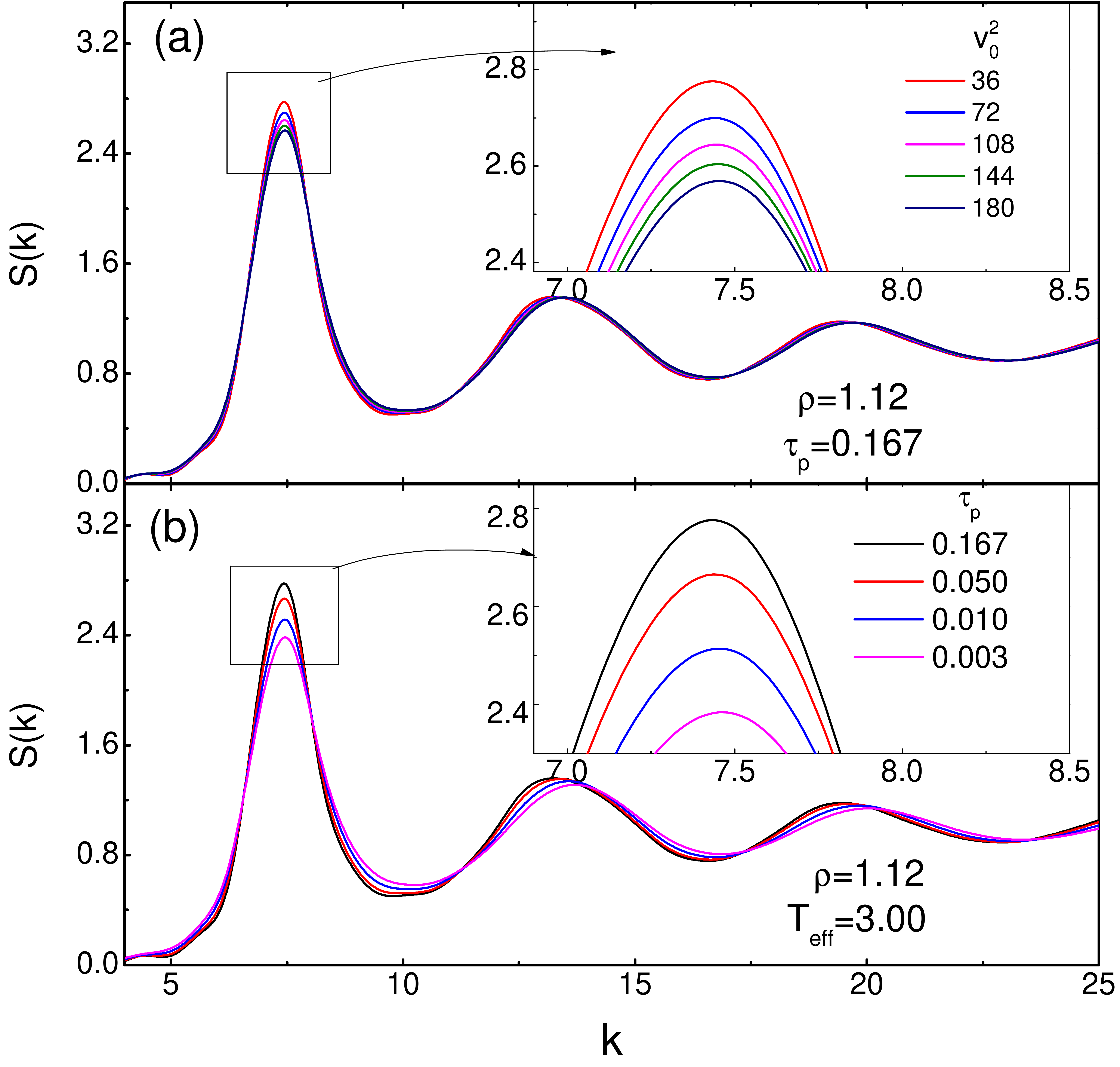}
\par\end{centering}
\caption{Non-equilibrium static structure factors $S\left(k\right)$for (a)
activity $v_{0}^{2}=$36 to 180 (in step of 36) with constant $\tau_{p}=0.167$,
(b) persistent time $\tau_{p}=$0.167, 0.05, 0.01, 0.003 with constant
$T_{{\rm eff}}=3.0$.}
\label{Fig_Sk}
\end{figure}

In Fig.\ref{Fig_Sk}(a), the non-equilibrium static structure factors
$S\left(k\right)$ obtained from direct simulations are drawn for
different particle activities $v_{0}$ for fixed $\rho=1.12$ and
$\tau_{p}=0.167$. The value of $\rho$ is chosen such that the system
is close to the glass transition and that of $\tau_{p}$ is consistent
with the setting in the simulation work of Ni\cite{2013_NC_NiRan_PushGlasTrans}.
It seems that $S\left(k\right)$ does not change much with the variation
of $v_{0}$, except that the main peak decreases slightly and shifts
a little to right with increasing $v_{0}$. This decreasing of the
main peak indicates that the structure becomes looser with increasing
active force. The other peaks at larger values of $k$ show little
discrepancy for different $v_{0}$. Such observations are in qualitative
agreements with the simulation results obtained by Ni. Since we have
fixed $\tau_{p}$, the effective temperature $T_{\text{eff}}\sim1+D_{f}\tau_{p}^{2}$
changes in the same tendency as $v_{0}$, such that Fig.\ref{Fig_Sk}(a)
also shows the change of $S\left(k\right)$ with $T_{\text{eff}}$.
In Fig.\ref{Fig_Sk}(b), $S\left(k\right)$ for different $\tau_{p}$,
but with fixed $T_{\text{eff}}$ have been presented. In this case,
one can see that the main peak increases apparently with increasing
$\tau_{p}$ and also shifts a little bit to smaller values of $k$.
Since $T_{\text{eff}}\sim1+v_{0}^{2}\tau_{p}/3$, increasing $\tau_{p}$
with fixed $T_{\text{eff}}$ corresponds to decreasing $v_{0}$, this
observation is consistent with Fig.\ref{Fig_Sk}(a). The second and
third peak also show observable differences with the variation of
$\tau_{p}$ in that the peak gets higher and moves to smaller values
of k with increasing $\tau_{p}$. 

\begin{figure}
\begin{centering}
\includegraphics[width=0.6\columnwidth]{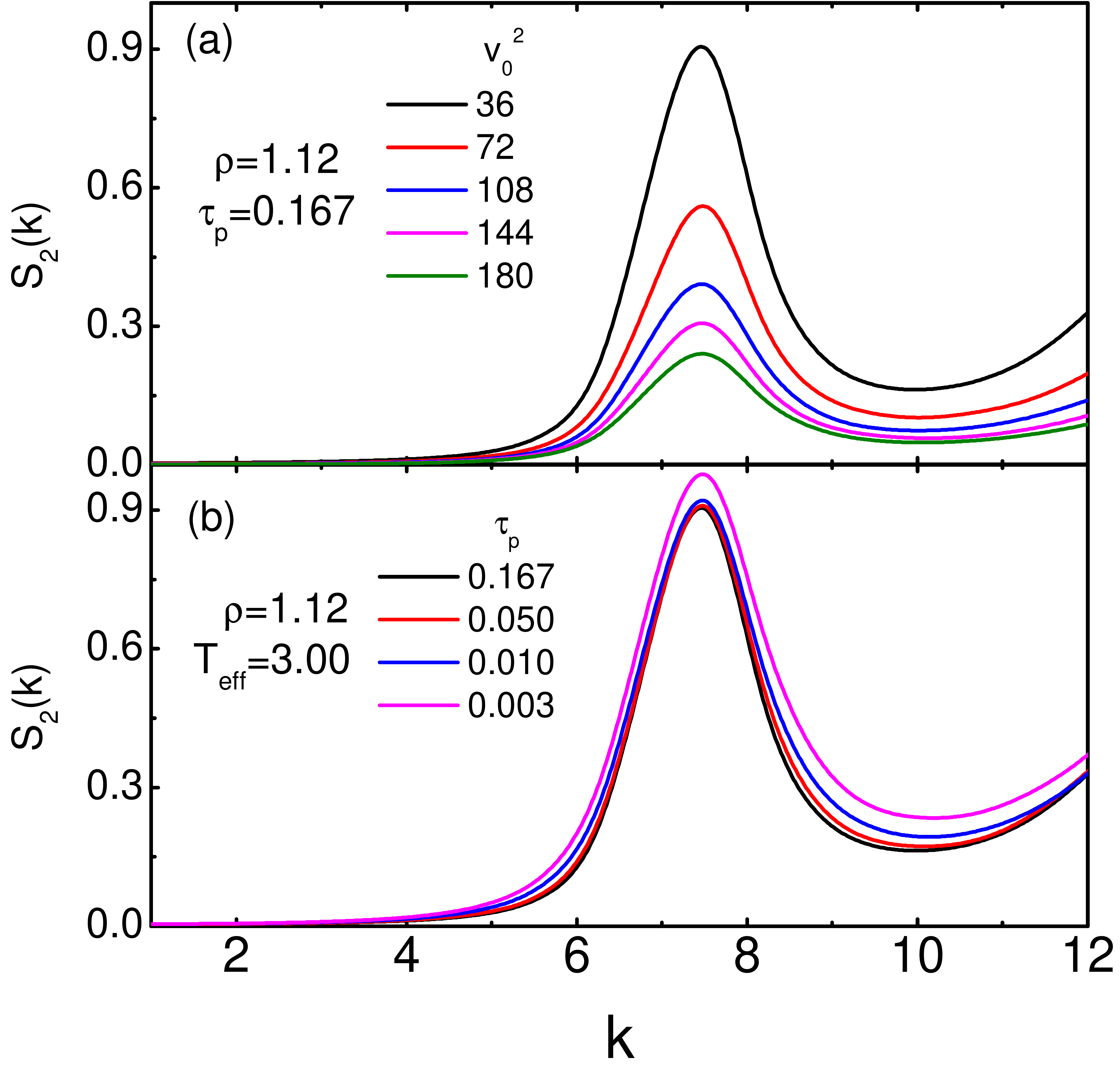}
\par\end{centering}
\caption{$S_{2}\left(k\right)$ for (a) activity $v_{0}^{2}=$36 to 180 (in
step of 36) with constant $\tau_{p}=0.167$, (b) persistent time $\tau_{p}=$0.167,
0.05, 0.01, 0.003 with constant $T_{{\rm eff}}=3.0$. }
\label{Fig_S2k}
\end{figure}

As discussed in the last section, an important new feature of our
theory is the introduction of the function $S_{2}\left(k\right)$,
which couples the instantaneous diffusion coefficient $D_{j}$ and
the density fluctuations. It is therefore instructive for us to investigate
how $S_{2}\left(k\right)$ looks like. In Fig.\ref{Fig_S2k}, we have
plotted $S_{2}\left(k\right)$ with the same parameter settings as
in Fig.\ref{Fig_Sk}. As shown in Fig.\ref{Fig_S2k}(a), the particle
activity (or effective temperature) drastically influences $S_{2}\left(k\right)$,
with the main peak decreasing considerably with increasing $v_{0}$
or $T_{\text{eff}}$. Compared to Fig.\ref{Fig_Sk}(a), the value
of $S_{2}\left(k\right)$ is much smaller than $S\left(k\right)$,
which reflects the fact that $D_{j}$ is generally less than $D_{0}$.
The behaviors of $S_{2}\left(k\right)$ for fixed $T_{\text{eff}}$
but with variant $\tau_{p}$ are shown in Fig.\ref{Fig_S2k}(b). In
this latter case, we see that the main peak now slightly reduces with
increasing $\tau_{p}$ and it seems to saturate for large $\tau_{p}$,
which are at variance with the observations in Fig.\ref{Fig_Sk}(b).
The apparent discrepancies between $S_{2}\left(k\right)$ and $S\left(k\right)$
indicate that our theory may show interesting new features.

\begin{figure}

\begin{centering}
n \includegraphics[width=0.6\columnwidth]{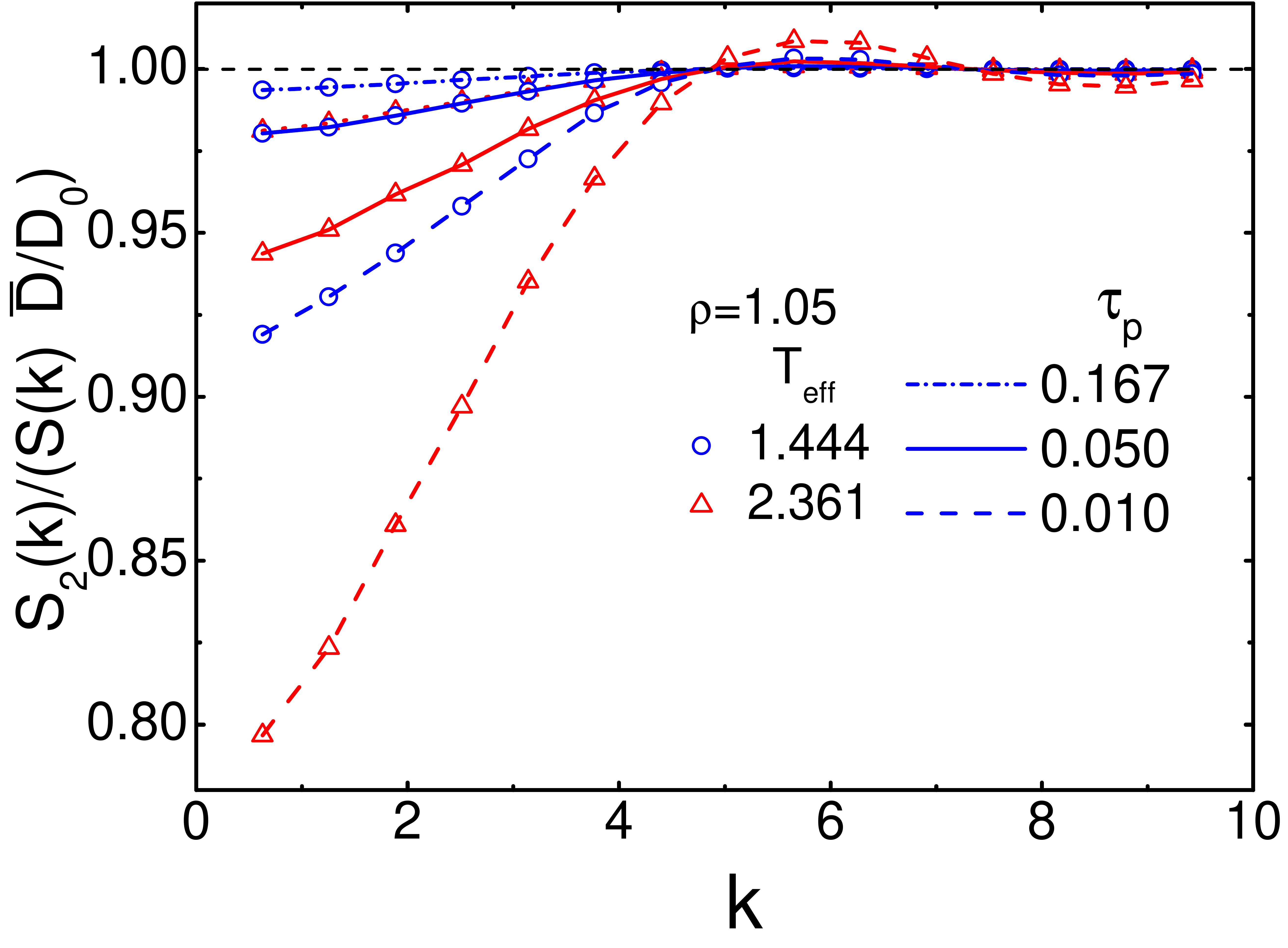}
\par\end{centering}
\caption{To show the difference between $S(k)$ and $D_{0}S_{2}\left(k\right)/\bar{D}$,
we plot $D_{0}S_{2}\left(k\right)/\bar{D}S(k)$ with different $T_{{\rm eff}}$=1.444,
2.361 and $\tau_{p}$= 0.010, 0.050, 0.167 , then find some structures
at small $k$. }

\label{fig-S2/SK}
\end{figure}

Another new feature of our theory is the pseudo-correlation function
$C_{2}\left(\mathbf{q},\mathbf{k}\right)$, which plays a similar
role to $c\left(k\right)$ in the irreducible memory function $M^{irr}\left(q,t\right)$.
As discussed above, $C_{2}\left(\mathbf{q},\mathbf{k}\right)$ reduces
to $c\left(k\right)$ if $S_{2}\left(p\right)D_{0}/\bar{D}S\left(p\right)\simeq1$.
In Fig.\ref{fig-S2/SK}, the dependence of $S_{2}\left(k\right)D_{0}/\bar{D}S\left(k\right)$
on $k$ has been presented, for fixed $\rho$ with varying $T_{{\rm eff}}$
and $\tau_{p}$. Interestingly, we find that it is approximately one
if $k$ is larger than $2\pi/\sigma$ which is approximately the peak
position for $S\left(k\right)$. Nevertheless, for small values of
$k$, $S_{2}\left(k\right)D_{0}/\bar{D}S\left(k\right)$ shows some
structures. Specifically, $S_{2}\left(k\right)D_{0}/\bar{D}S\left(k\right)$
becomes much less than one for small $\tau_{p}$ if $T_{\text{eff}}$
is fixed. Such a feature may lead to enhancement of the irreducible
memory function $M_{s}^{irr}\left(q,t\right)$ shown in Eq.(\ref{eq:M_irr_s})
with decreasing $\tau_{p}\to0$ if $T_{\text{eff}}$ is fixed. This
would lead to the increment of $\tau_{\alpha}$,if other effects
are not accounted for. Note that, however, $\bar{D}$ increases with
decreasing $\tau_{p}$ with constant $T_{\text{eff}}$ as shown in
Fig.\ref{Fig_Dbar}, such that $\tau_{\alpha}$ would decrease with
decreasing $\tau_{p}$ in terms of this effect. Therefore, it might
be possible that the relaxation time $\tau_{\alpha}$ shows some re-entrance
behaviors in the small $\tau_{p}$ region, similar to that reported
for the AOU model\cite{2014_PRL_Berthier_GTActiveHardDisk,2015_PRE_Szamel_GlasDyna}. 

\subsection{Intermediate Scattering Function }

\begin{figure}
\centering{}\includegraphics[width=0.8\columnwidth]{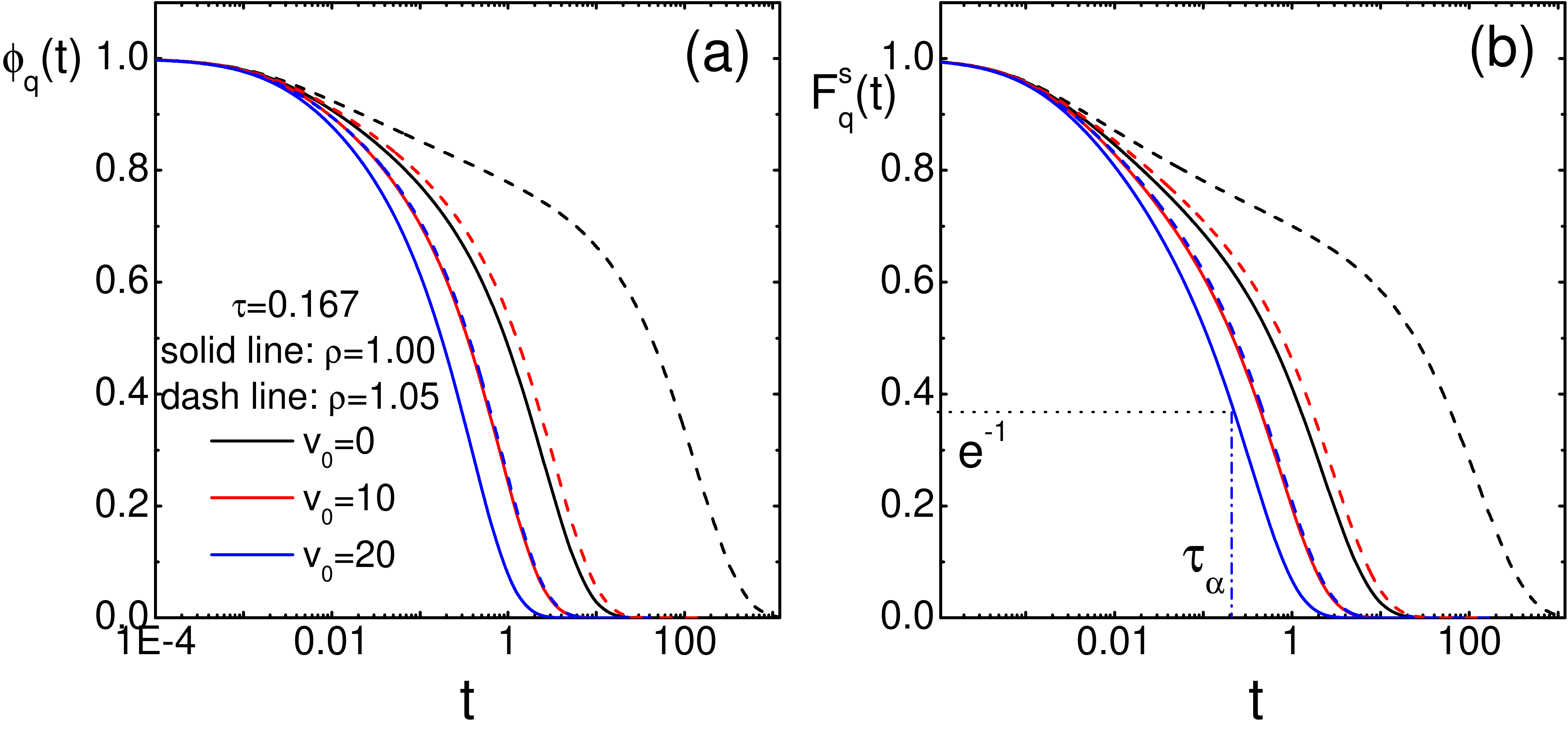}\caption{Intermediate scattering function $\phi_{q}\left(t\right)$ in (a)
and Self-intermediate scattering function $F_{q}^{s}\left(t\right)$
in (b), change with $q=7.5$, density $\rho=1.00$, 1.05 and activity
$v_{0}$=0, 10, 20. As well as the definition of relaxation time $\tau_{\alpha}$.
Notice that the same color and shape of lines in (a) and (b) denotes
the same parameter.}
\label{Fig_Fq_t}
\end{figure}

With the static properties obtained above, particularly $\bar{D}$
and $S_{2}\left(k\right)$, we are ready to investigate the behavior
of the intermediate scattering function $F_{q}\left(t\right)$ by
numerically solving the memory functions Eqs. (\ref{eq:GLE_Fq}) and
(\ref{eq:GLE_Fsqt}). In Fig.\ref{Fig_Fq_t}(a), the normalized scattering
functions $\phi_{q}\left(t\right)=F_{q}\left(t\right)/S\left(q\right)$
are shown for different values of $v_{0}$ (or $T_{\text{eff}}$)
and number density $\rho$, wherein we have chosen $q=7.5$ which
is around the first peak of $S\left(q\right)$. The results for two
densities $\rho=1.05$ and 1.10 are plotted, and the value of $\tau_{p}$
is fixed to be 0.167. For the higher density $\rho=1.07,$ one can
see that $F_{q}\left(t\right)$ finally reaches a plateau in the long
time limit for $v_{0}=0$ (or $T_{\text{eff}}=1$), indicating that
the system reaches the glassy state. For a nonzero value of $v_{0}$
as shown in the figure, $F_{q}\left(t\right)$ will finally relax
to zero for large $t$ indicating that the system is in a liquid state,
and the relaxation time decreases apparently with increasing $v_{0}$.
Therefore, activity will push the glass transition to higher number
density, in consistent with the simulation results of the RAB model
and other related models. For a smaller $\rho=1.05$, the system is
in the liquid state for $v_{0}=0$ and the relaxation of $F_{q}\left(t\right)$
also becomes faster with increasing $v_{0}$ or $T_{\text{eff}}$.
The behaviors of the self-scattering function $F_{q}^{s}\left(t\right)$
are similar as shown in Fig.\ref{Fig_Fq_t}(b). While for $v_{0}=0$
the tracer particle is trapped and $F_{q}^{s}\left(t\right)$ reaches
a non-zero value for $t\to\infty$, it relaxes to zero for $v_{0}=10$
and $20$ with the relaxation time $\tau_{\alpha}$ decreases apparently
with increasing $v_{0}$. 

The limiting value $f_{q}=\lim_{t\to\infty}\phi_{q}\left(t\right)$
at the plateau defines the so-called Debye-Waller factor. A non-zero
value of $f_{q}$ indicates that the system is in the glassy state.
With the increase of $\rho$, $f_{q}$ may change from zero to an
apparent nonzero value, and the value $\rho_{c}$ thus corresponds
to the glass transition point. One may also fix $\rho$ but vary $T_{\text{eff}}$,
then $f_{q}$ may become nonzero for $T_{\text{eff}}$ less than some
critical value $T_{\text{eff}}^{c}$ , which defines a critical temperature
for glass transition. According to the MCT Eq. (\ref{eq:GLE_Fq}),
the Debye-Waller factor $f_{q}$ follows
\begin{equation}
f_{q}=\frac{m_{q}(\infty)}{1+m_{q}(\infty)}\label{eq:DWP}
\end{equation}
where 
\begin{equation}
{\normalcolor m_{q}(\infty)=\frac{\rho\bar{D}}{16\pi^{3}q^{2}}\int d^{3}\mathbf{k}\left[\left(\hat{\mathbf{q}}\cdot\mathbf{k}\right)C_{2}\left(\mathbf{q};\mathbf{k}\right)+\left(\hat{\mathbf{q}}\cdot\mathbf{p}\right)C_{2}(\mathbf{q};\mathbf{p})\right]^{2}S(k)S(q)S(p)f_{k}f_{p}}
\end{equation}
This equation can be solved numerically and self-consistently to get
$f_{q}$ for given control parameters $v_{0}$ (or $T_{\text{eff}}$),
$\tau_{p}$ as well as $\rho$.

\begin{figure}
\centering{}\includegraphics[width=0.8\columnwidth]{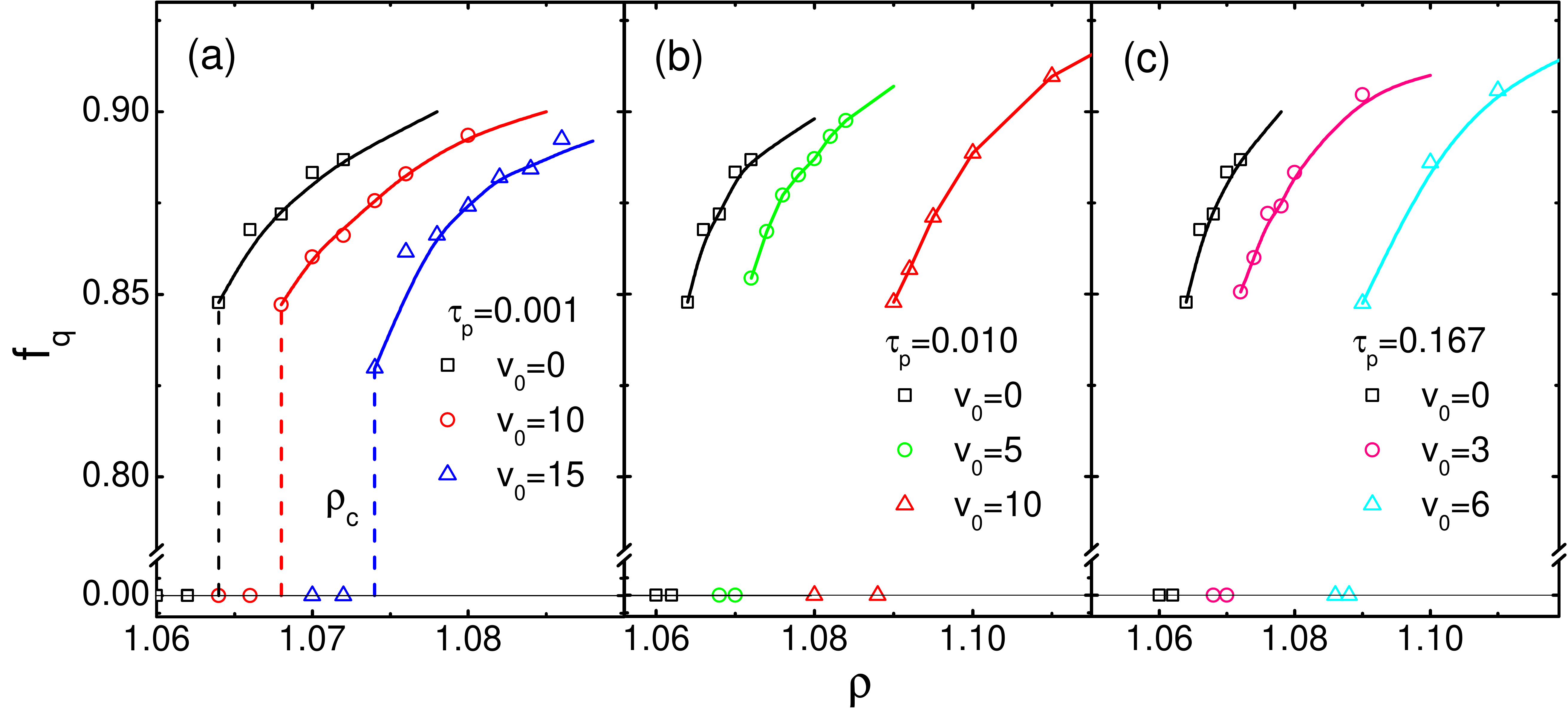}\caption{Debye-Waller factor as a function of density $\rho$, for different
$v_{0}$ (as well$T_{\text{eff}}$) and $\tau_{p}$=0.001 in (a),
0.010 in (b), and 0.167 in (c), with different $v_{0}$.}
\label{Fig_fq_rho}
\end{figure}

In Fig.\ref{Fig_fq_rho}(a), the dependence of $f_{q}$ on the number
density $\rho$ is presented for different $v_{0}$ (or $T_{\text{eff}}$)
with given $\tau_{p}=0.001$. Clearly, $f_{q}$ changes abruptly from
zero to a large nonzero value at a critical density $\rho_{c}$, indicated
for example by the vertical dashed line for $v_{0}=0$ at about $\rho_{c}\simeq1.064$.
For given $\tau_{p}$, the curve shifts to larger values of $\rho$
with increasing $v_{0}$, indicating that that glass transition is
pushed to higher values of $\rho$ for larger particle activity in
consistent with previously reported simulation results. The pictures
for different values of $\tau_{p}=0.05$ and 0.167 are shown in Fig.\ref{Fig_fq_rho}(b)
and (c), respectively. The results are similar to those in (a), with
the values of $\rho_{c}$ shifting to relatively larger values for
larger $\tau_{p}$. 

\begin{figure}
\begin{centering}
\includegraphics[width=0.9\columnwidth]{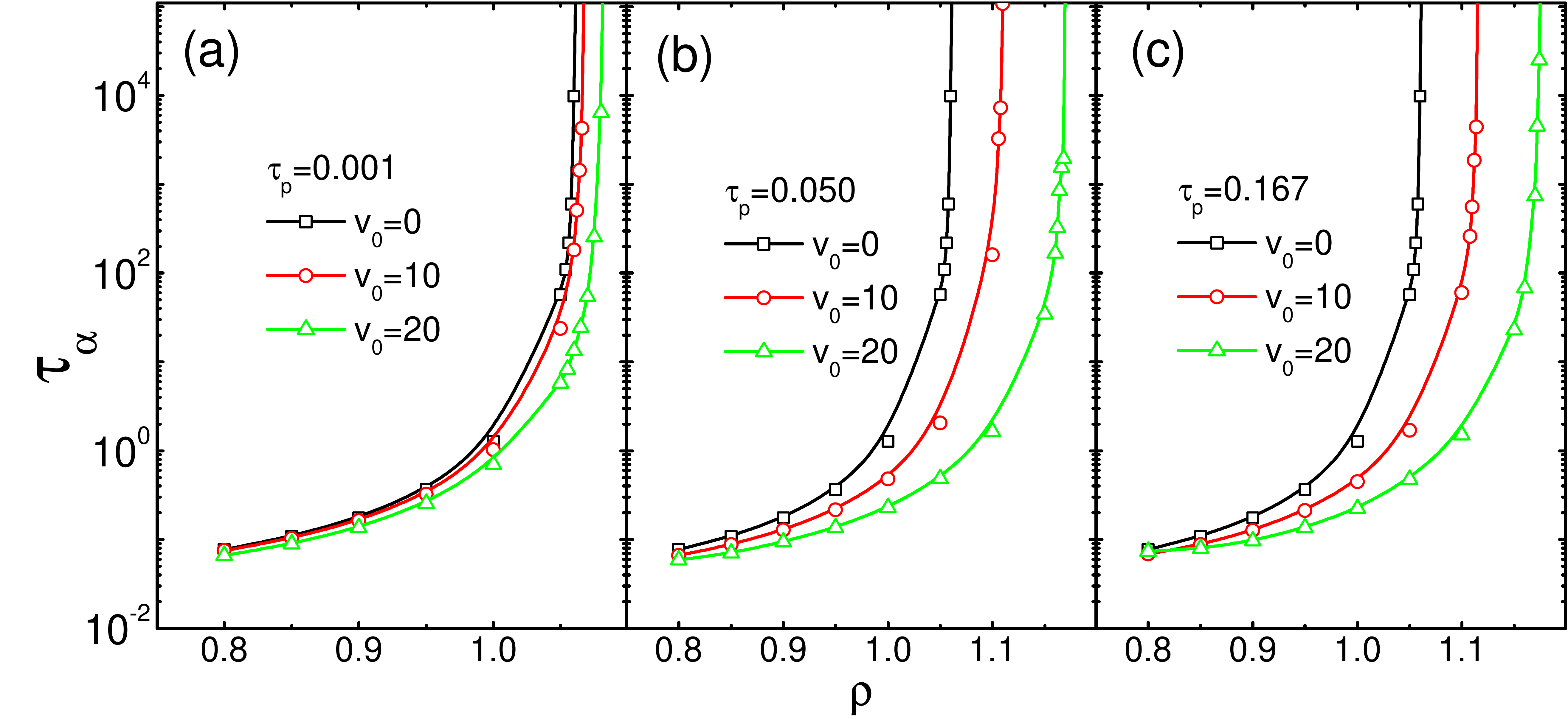}
\par\end{centering}
\caption{Relaxation time $\tau_{\alpha}$ as a function of density $\rho$,
for different $v_{0}$ (as well$T_{\text{eff}}$) and $\tau_{p}$=0.001
in (a), 0.010 in (b), and 0.167 in (c), with $v_{0}=$0, 10, 20.}
\label{Fig_tau_rho}
\end{figure}

In Fig.\ref{Fig_Fq_t}(b), it is shown that the relaxation time $\tau_{\alpha}$
increases when the system approaches the glass transition and it diverges
at the glass transition point. Therefore, one may also study the glass
transition by investigating the behavior of $\tau_{\alpha}$ as a
function of $\rho$. The results are depicted in Fig.\ref{Fig_tau_rho}
with the same parameter setting as in Fig.\ref{Fig_fq_rho}. Obviously,
$\tau_{\alpha}$ increases fastly with $\rho$ for fixed values of
$v_{0}$ ($T_{\text{eff}}$) and $\tau_{p}$ and it diverges at some
critical value $\rho_{c}$. For a very small $\tau_{p}=0.001$, it
seems that changing $v_{0}$ does not affect very much the values
of $\tau_{\alpha}$ as shown in Fig.\ref{Fig_tau_rho}(a). The influence
becomes more considerable when $\tau_{p}$ gets larger as demonstrated
in \ref{Fig_tau_rho}(b) and (c), and the value of $\rho_{c}$ also
shifts to larger values in consistent with Fig.\ref{Fig_fq_rho}. 

\begin{figure}
\begin{centering}
\par\end{centering}
\begin{centering}
\includegraphics[width=0.8\columnwidth]{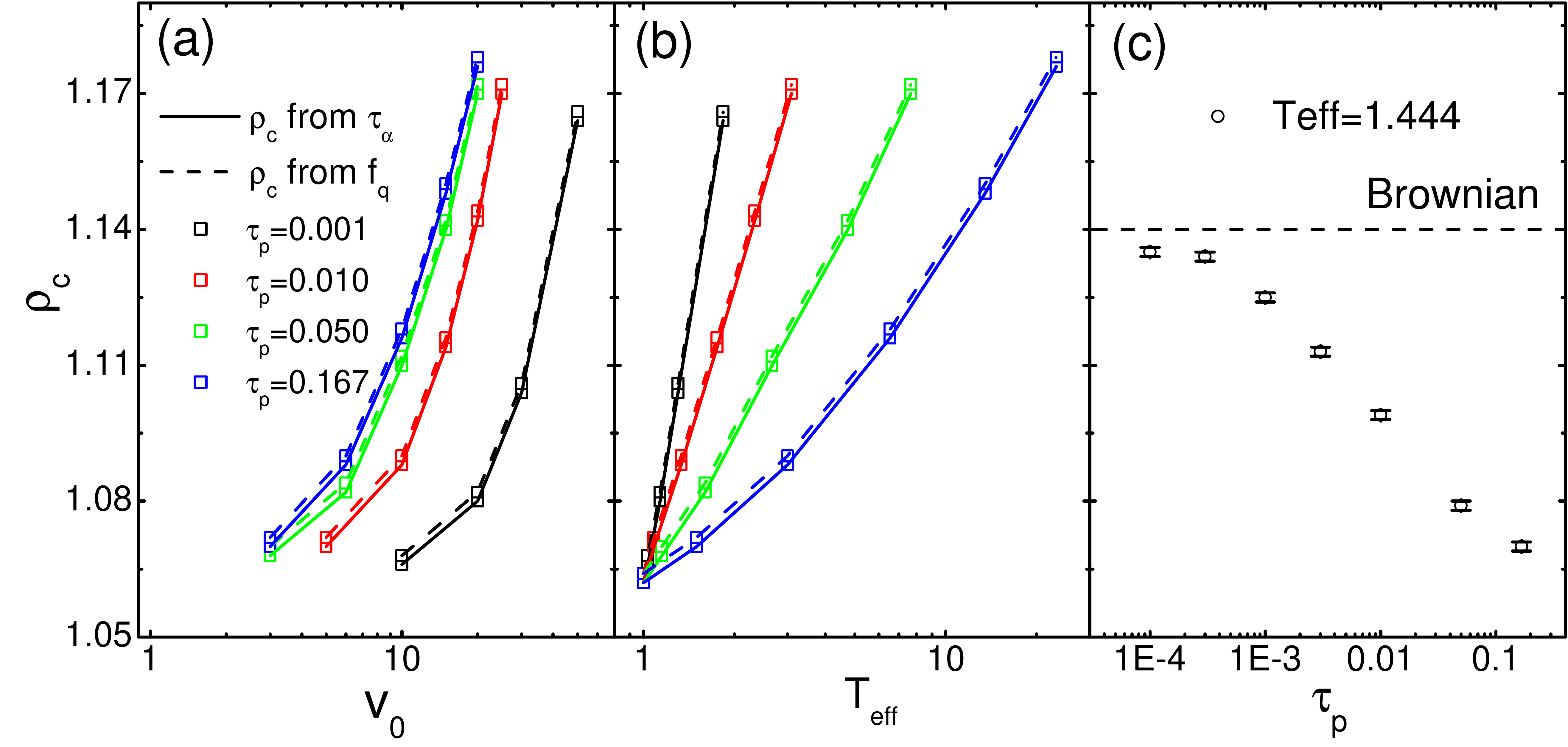}
\par\end{centering}
\begin{centering}
\par\end{centering}
\caption{MCT predict critical density $\rho_{c}$ as function of $v_{0}$ in
(a) and $T_{{\rm eff}}$ in (b), with $\tau_{p}$=0.001(black), 0.010(red),
0.050(green), and 0.167(blue). The solid and dash lines mean different
methods to get $\rho_{c}$. In figure(c) we also show the dependence
of $\rho_{c}$ on $\tau_{p}$, with the constant effective temperature
$T_{{\rm eff}}=1.444$. }
\label{Fig: Rho_c}
\end{figure}

Surely, the value of $\rho_{c}$ should be the same either obtained
by $f_{q}$ or $\tau_{\alpha}$ within reasonable fluctuations. In
Fig.\ref{Fig: Rho_c}(a) and (b), the dependence of $\rho_{c}$, obtained
from both$f_{q}$ and $\tau_{\alpha}$, on $v_{0}$ and $T_{\text{eff }}$
are presented for different given values of $\tau_{p}$. Clearly,
$\rho_{c}$ increases with both $v_{0}$ and $T_{\text{eff}}$ as
expected. Interestingly, $\rho_{c}$ shows a nearly linear dependence
on $v_{0}^{2}$, wherein the slope increases with $\tau_{p}$. This
linear dependence was also observed in the simulation work of Ni.
We also note that $\rho_{c}$ increases with $\tau_{p}$ for fixed
$v_{0}$, whereas it decreases with $\tau_{p}$ for fixed $T_{\text{eff}}$
according to the data presented in Fig.\ref{Fig: Rho_c}(a) and (b).
This is shown more clearly in Fig.\ref{Fig: Rho_c}(c) , where we
have also plotted $\rho_{c}$ as a function of $\tau_{p}$ for different
$T_{\text{eff}}$. For comparison, the dashed line gives the value
of $\rho_{c}^{B}$ for the corresponding passive Brownian system with
$T=T_{\text{eff}}$, which is obtained by setting $D_{t}=D_{0}$ and
zero self-propulsion force $f_{i}=0$ in Eq.(\ref{eq:AOU-T-1}). Clearly,
$\rho_{c}$ approaches $\rho_{c}^{B}$ in the limit $\tau_{p}\to0$
as expected. 

\begin{figure}

\begin{centering}
\includegraphics[width=0.8\linewidth]{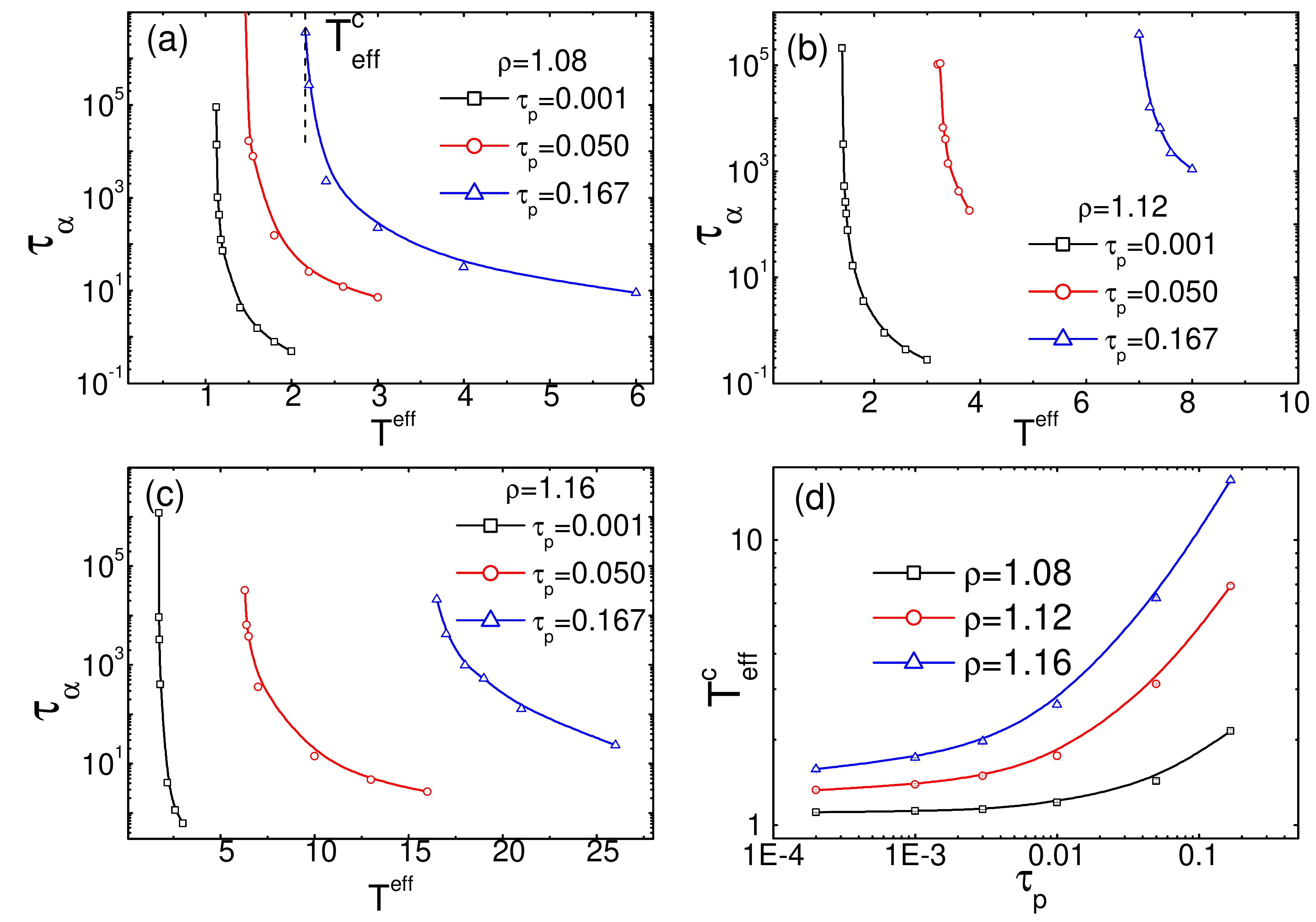}\caption{The relaxation time $\tau_{\alpha}$ as a function of $T_{{\rm eff}}$,
for $\tau_{p}$= 0.001, 0.050, 0.167 with density $\rho$=1.08 in
(a), 1.12 in (b), and 1.16 in (c). And according to these data, we
show the dependence of MCT predicting critical effective temperature
$T_{{\rm eff}}^{c}$ on $\tau_{p}$ for different densities in figure(d). }
\label{fig: tau_a-Teff}
\par\end{centering}
\end{figure}
For equilibrium systems, glass transition is often studied in terms
of the critical temperature $T_{c}$, below which the system enters
the glassy state. In the present work, we may study the non-equilibrium
glass transition in the same spirit by calculating the value of critical
effective temperature $T_{\text{eff}}^{c}$ with fixed number density
$\rho$. In Fig.\ref{fig: tau_a-Teff} (a) to (c), the results of
$\tau_{\alpha}$ are presented as functions of $T_{\text{eff}}$ for
different $\tau_{p}$ and $\rho$. For fixed value of $\tau_{p}$
and $\rho$, $\tau_{\alpha}$ decreases monotonically with $T_{\text{eff}}$.
Below some critical values of $T_{\text{eff}}$ corresponding to $T_{\text{eff}}^{c}$,
$\tau_{\alpha}$ diverges indicating the occurrence of glass transition.
In Fig.\ref{fig: tau_a-Teff} (d), the dependence of $T_{\text{eff}}^{c}$
on $\tau_{p}$ for different $\rho$ is shown. One can see that $T_{\text{eff}}^{c}$
increases monotonically with $\tau_{p}$, and it approaches a constant
value in the limit $\tau_{p}\to0$. Such a constant value corresponds
to the one for a passive Brownian system with $T_{c}=T_{\text{eff}}^{c}$.
We also note that $T_{\text{eff}}^{c}$ increases with the number
density $\rho$, indicating that a denser system enters glass transition
at a higher critical temperature as expected.

\section{Conclusions\label{sec:Conclusions}}

In summary, we have developed a promising mode coupling theory to
study the nonequilibrium glassy dynamics of a general model system
of self-propelled particles. The self-propulsion force is given by
a colored noise described by OU process, and thermal noises in the
environment are also considered. Our work mainly contains two parts.
By using Fox approximation method for Langevin systems with colored
noise, an approximate Smoluchowski equation can be obtained, governing
the time evolution of the distribution function of the particles'
positions. This effective SE is expected to be exactly valid for not
large persistence time $\tau_{p}$ of the propulsion force, and it
thus serves as a promising starting point to study the system's relaxation
or glassy dynamics. The SE involves a configuration dependent instantaneous
diffusion function $D_{j}\left(\mathbf{r}^{N}\right)$ which is related
to the gradient of force subjected to particle $j$. With this SE,
we are able to derive a memory function equation for the time dependent
behavior of the collective or self- intermediate scattering function
$F_{q}\left(t\right)$ or $F_{q}^{s}\left(t\right)$ in the nonequilibrium
steady state. Applying the basic assumption that macroscopic currents
vanish in the steady state and using standard mode coupling approximation,
we have obtained the expressions for the irreducible memory function
as well as frequency terms. Particularly, we find that the dynamics
are mainly determined by an effective diffusion coefficient $\bar{D}$,
which is the ensemble average of $D_{j}\left(\mathbf{r}_{N}\right)$
in the nonequilibrium steady state, and a pseudo steady state structure
factor $S_{2}\left(k\right)$, which involves the coupling between
$D_{j}\left(\mathbf{r}^{N}\right)$ and density fluctuations. $\bar{D}$
enters the frequency term and thus governs the short time dynamics,
whereas both enter the vortex for memory function and influence the
long time dynamics. By direct simulations, we find that $\bar{D}$
increases with the single effective temperature $T_{\text{eff}}$
as well as the magnitude $v_{0}$ of propulsion force, while it decreases
with $\tau_{p}$ for fixed $T_{\text{eff}}$ or $v_{0}$. The structure
function $S_{2}\left(k\right)$ simply decouples into the product
of $\bar{D}/D_{0}$ and $S\left(k\right)$, with $S\left(k\right)$
the nonequilibrium static structure factor and $D_{0}$ the single
particle diffusion coefficient in the limit $\tau_{p}\to0$, for relatively
large values of $k$, whereas it shows apparent deviations from $\left(\bar{D}/D_{0}\right)\cdot S\left(k\right)$
for small $k$s. Our theory makes it feasible for us to investigate
the glassy dynamics of the system, by investigating the time behavior
of $F_{q}\left(t\right)$ or $F_{q}^{s}\left(t\right)$ in the long
time limit, chosen the persistence time $\tau_{p}$, the effective
temperature $T_{\text{eff}}$, as well as the number density $\rho$
as free parameters. We find that the critical density $\rho_{c}$
for glass transition shifts to larger values with increasing $T_{\text{eff}}$
or $v_{0}$ if $\tau_{p}$ is fixed, in good qualitative accordance
with the simulation results of active Brownian particles and related
systems. In addition, we have also investigated how the critical density
$\rho_{c}$ changes with $\tau_{p}$ for a fixed $T_{\text{eff}}$,
finding that $\rho_{c}$ decreases with $\tau_{p}$ monotonically
and it approaches the value for the corresponding passive Brownian
system in the limit $\tau_{p}\to0$ as expected. We have also calculated
the critical temperature $T_{\text{eff}}^{c}$ for glass transition
at fixed density, finding that it increases monotonically with $\tau_{p}$
and also approaches the Brownian particle limit for $\tau_{p}\to0$. 

In future work, we would like to extend the present method to more
complex systems such as mixtures of self-propelled particles but with
different driving forces or to mixtures of active-passive particles\cite{2015_our_Arxiv}.
As mentioned in the main text, the relaxation time $\tau_{\alpha}$
may show nontrivial dependence on the persistence time $\tau_{p}$,
which would be also an interesting topic to address for system with
both propulsion force and thermal noise. In addition, our results
demonstrate that only in the limit $\tau_{p}\to0$, the glass transition
point $\rho_{c}$ or $T_{\text{eff}}^{c}$ approaches that of a Brownian
system, indicating that the 'collective' effective temperature with
respect to the nonequilibrium glass transition is different from the
single particle one \cite{2015_EPL_Levis_Single-collective-effecTemp},
which may deserve more detailed study. In a word, we believe that
our work presents a useful theoretical framework to study the nonequilibrium
dynamics of dense active particles system from the microscopic level
which could find many applications in future works. 
\begin{acknowledgments}
This work is supported by National Basic Research Program of China(Grant
No. 2013CB834606), by National Science Foundation of China (Grant
Nos. 21673212, 21521001, 21473165, 21403204), by the Ministry of Science
and Technology of China (Grant No.�2016YFA0400904), and by the Fundamental
Research Funds for the Central Universities (Grant Nos. WK2060030018,
2030020028,2340000074). 
\end{acknowledgments}

\numberwithin{equation}{section} 

\appendix

\section{Derivation of the Smoluchowski Equation\label{sec:a}}

Generally, for a LE with colored noise, one can get the FPE within
Fox approximation\cite{1986_PRA_Fox}. For illustration, consider
a simple one dimensional over damped LE 
\begin{equation}
\dot{x}\left(t\right)=G\left(x\right)+\chi\left(t\right)
\end{equation}
where $G(x)$ denotes the external or internal force and $\chi\left(t\right)$
is the stochastic noise with correlation 
\begin{equation}
\left\langle \chi\left(t\right)\chi\left(s\right)\right\rangle =C\left(t-s\right)
\end{equation}
Define a probability distribution function 
\begin{equation}
P\left(y,t\right)=\int D\left[\chi\right]P\left[\chi\right]\delta\left(y-x\left(t\right)\right)
\end{equation}
where $D\left[\chi\right]$ denotes integration over the noisy path
of $\chi\left(t\right)$ and $P\left[\chi\right]$ is the distribution
functional of $\chi$ which is assumed to be Gaussian. One can then
obtain the FPE governing the evolution of $P\left(y,t\right)$ as
follows \cite{2015_PRE_Farage_EffectIntera}
\begin{equation}
\frac{\partial}{\partial t}P\left(y,t\right)=-\frac{\partial}{\partial y}\left[G\left(y\right)P\left(y,t\right)\right]+\frac{\partial^{2}}{\partial y^{2}}\left\{ \int_{0}^{t}ds'C\left(t-s'\right)\int D\left[\chi\right]P\left[\chi\right]e^{\int_{s'}^{t}dsG'\left(x\left(s\right)\right)}\delta\left(y-x\left(t\right)\right)\right\} 
\end{equation}
Note that if $\chi\left(t\right)$ is white noise, $C\left(t-s\right)=D_{0}\delta\left(t-s\right)$,
then the second term is just 
\begin{equation}
D_{0}\frac{\partial^{2}}{\partial y^{2}}\left[\int D\left[\chi\right]P\left[\chi\right]\delta\left(y-x\left(t\right)\right)\right]\equiv D_{0}\frac{\partial^{2}}{\partial y^{2}}P\left(y,t\right)
\end{equation}
which recovers the standard FPE. For a colored noise with 
\begin{equation}
C\left(t-s\right)=\frac{D}{\tau_{0}}\exp\left(-\frac{\left|t-s\right|}{\tau_{0}}\right)
\end{equation}
one can obtain the FPE approximately as
\begin{equation}
\frac{\partial}{\partial t}P\left(y,t\right)=-\frac{\partial}{\partial y}\left[G\left(y\right)P\left(y,t\right)\right]+D\frac{\partial^{2}}{\partial y^{2}}\left[\frac{1}{1-\tau_{0}G'\left(y\right)}P\left(y,t\right)\right]
\end{equation}
by assuming 
\begin{equation}
\int_{s'}^{t}dsG'\left(x\left(s\right)\right)\approx G\left(x\left(t\right)\right)\left(t-s'\right)
\end{equation}

For a general multi-variable case, 
\begin{equation}
\frac{dx_{i}\left(t\right)}{dt}=G_{i}\left(\{x_{i}\}\right)+\chi_{i}\left(t\right)
\end{equation}
where
\begin{equation}
\left\langle \chi_{i}\left(t\right)\chi_{j}\left(s\right)\right\rangle =C_{ij}\left(t-s\right)
\end{equation}
with $i,j=1,2,\cdots,N$, the FPE for distribution function 
\begin{equation}
P\left(\mathbf{y},t\right)=\int D\left[\chi\right]P\left[\chi\right]\delta\left(\mathbf{y}-\mathbf{x}\left(t\right)\right)
\end{equation}
reads
\begin{eqnarray*}
\frac{\partial}{\partial t}P\left(\mathbf{y},t\right) & = & -\sum_{i}\partial_{i}\left[G_{i}\left(\mathbf{y}\right)P\left(\mathbf{y},t\right)\right]\\
 &  & +\sum_{ij}\partial_{i}\left\{ \sum_{l}\int_{0}^{t}ds'C_{il}\left(t-s'\right)\partial_{j}\int D\left[\chi\right]P\left[\chi\right]\exp\left[\int_{s'}^{t}ds\frac{\partial}{\partial x_{l}}G_{j}\left(\mathbf{x}\left(s\right)\right)\delta_{jl}\right]\delta\left(\mathbf{y}-\mathbf{x}\left(t\right)\right)\right\} \\
 & = & -\sum_{i}\partial_{i}\left[G_{i}\left(\mathbf{y}\right)P\left(\mathbf{y},t\right)\right]\\
 &  & +\sum_{ij}\partial_{i}\left\{ \int_{0}^{t}ds'C_{ij}\left(t-s'\right)\partial_{j}\int D\left[\chi\right]P\left[\chi\right]\exp\left[\int_{s'}^{t}ds\partial{}_{j}G_{j}\left(x\left(s\right)\right)\right]\delta\left(\mathbf{y}-\mathbf{x}\left(t\right)\right)\right\} 
\end{eqnarray*}

Then, if $C_{ij}\left(t-s\right)=\delta_{ij}C\left(t-s\right)=\delta_{ij}\frac{D}{\tau_{0}}\exp\left(-\frac{\left|t-s\right|}{\tau_{0}}\right)$,
using the assumption mentioned before we can get
\begin{equation}
\frac{\partial}{\partial t}P\left(\mathbf{y},t\right)=-\sum_{i}\partial_{i}\left[G_{i}\left(\mathbf{y}\right)P\left(\mathbf{y},t\right)\right]+\sum_{i}D\partial_{i}^{2}\left\{ \left[\frac{1}{1-\tau_{0}\partial_{i}G_{i}\left(\mathbf{y}\right)}\right]P\left(\mathbf{y},t\right)\right\} \label{eq:A12}
\end{equation}

For our system described by Eq.(\ref{eq:AOU-T-1}), we have correspondingly
$x\to\mathbf{r}^{N}$, $G\left(x\right)\to\gamma^{-1}\mathbf{F}\left(\mathbf{r}^{N}\right)=\beta D_{t}\mathbf{F}\left(\mathbf{r}^{N}\right)$,
$\gamma^{-1}\mathbf{f}_{i}\left(t\right)\to\mathbf{\chi}_{i}\left(t\right)$.
According to Eq.(\ref{eq:fi_corr}), the variable $D$ in Eq. (\ref{eq:A12})
is $D_{f}\tau_{p}^{2}/\gamma^{2}$ and $\tau_{0}$ is $\tau_{p}$.
Note that the white noise term $\mathbf{\eta}_{i}\left(t\right)$
in Eq.(\ref{eq:AOU-T-1}) contributes a normal diffusion term to the
FPE. Thus we finally obtain
\begin{eqnarray*}
\frac{\partial}{\partial t}\Psi\left(\mathbf{r}^{N},t\right) & = & -\sum_{i}\nabla_{i}\cdot\left[\beta D_{t}\mathbf{F}_{i}\left(\mathbf{r}^{N}\right)-D_{t}\nabla_{i}\right]\Psi\left(\mathbf{r}^{N},t\right)\\
 &  & +\sum_{i}\nabla_{i}^{2}\left[\frac{D_{f}\tau_{p}^{2}/\gamma^{2}}{1-\tau_{p}\cdot\beta D_{t}\nabla_{i}\mathbf{F}_{i}\left(\mathbf{r}^{N},t\right)}\right]\Psi\left(\mathbf{r}^{N},t\right)\\
 & = & +\sum_{i}\nabla_{i}\cdot\left\{ D_{t}+\left[\frac{D_{f}\tau_{p}^{2}/\gamma^{2}}{1-\tau_{p}\cdot\beta D_{t}\nabla_{i}\mathbf{F}_{i}\left(\mathbf{r}^{N}\right)}\right]\right\} \cdot\nabla_{i}\Psi\left(\mathbf{r}^{N},t\right)\\
 &  & -\sum_{i}\nabla_{i}\cdot\left\{ \beta D_{t}\mathbf{F}_{i}\left(\mathbf{r}^{N}\right)-\nabla_{i}\left[\frac{D_{f}\tau_{p}^{2}/\gamma^{2}}{1-\tau_{p}\cdot\beta D_{t}\nabla_{i}\mathbf{F}_{i}\left(\mathbf{r}^{N}\right)}\right]\right\} \Psi\left(\mathbf{r}^{N},t\right)
\end{eqnarray*}
Write
\begin{equation}
D_{i}\left(\mathbf{r}^{N}\right)=D_{t}+\left[\frac{D_{f}\tau_{p}^{2}/\gamma^{2}}{1-\tau_{p}\cdot\beta D_{t}\nabla_{i}\mathbf{F}_{i}\left(\mathbf{r}^{N}\right)}\right]
\end{equation}
and 
\begin{eqnarray}
\mathbf{F}_{i}^{\mbox{eff}}\left(\mathbf{r}^{N}\right) & = & \frac{D_{t}}{D_{i}\left(\mathbf{r}^{N}\right)}\left\{ \mathbf{F}_{i}\left(\mathbf{r}^{N}\right)-\frac{1}{\beta D_{t}}\nabla_{i}\left[\frac{D_{f}\tau_{p}^{2}/\gamma^{2}}{1-\tau_{p}\cdot\beta D_{t}\nabla_{i}\mathbf{F}_{i}\left(\mathbf{r}^{N}\right)}\right]\right\} \nonumber \\
 & = & \frac{D_{t}}{D_{i}\left(\mathbf{r}^{N}\right)}\left[\mathbf{F}_{i}\left(\mathbf{r}^{N}\right)-\frac{1}{\beta D_{t}}\nabla_{i}D_{i}\left(\mathbf{r}^{N}\right)\right]
\end{eqnarray}
is the effective force. Finally, we have 
\begin{equation}
\frac{\partial}{\partial t}\Psi\left(\mathbf{r}^{N},t\right)=-\sum_{i}\nabla_{i}\cdot D_{i}\left(\mathbf{r}^{N}\right)\cdot\left[\nabla_{i}-\beta\mathbf{F}_{i}^{\mbox{eff}}\left(\mathbf{r}^{N}\right)\right]\Psi\left(\mathbf{r}^{N},t\right)
\end{equation}
which corresponds exactly to Eqs. (\ref{eq:SmoFunc}) to (\ref{eq:F_eff})
in the main text. 

\section{Derivation of the General Langevin Equation\label{appGLE}}

\subsection{Memory Function Equation }

Here we present the derivation of the memory function equations, namely
Eqs. (\ref{eq:GLE_Fq}) to (\ref{eq:S2_k}) in the main text, for
the scattering function $F_{q}\left(t\right)=\frac{1}{N}\left\langle \text{\ensuremath{\rho}}_{\mathbf{q}}^{*}e^{\hat{\Omega}t}\rho_{\mathbf{q}}\right\rangle $.
This is most easily done in the Laplace domain, even for the complex
Smoluchowski operator $\hat{\Omega}$ shown in Eq.(\ref{eq:SmoOpe})
which contains the instantaneous effective diffusion constant $D_{j}\left(\mathbf{r}^{N}\right)$
given by Eq.(\ref{eq:D_j}). The main steps are similar to the derivation
of MCT equations for passive colloidal systems, following Mori-Zwanzig
projection operator procedures. 

We start from Laplace transform of the scattering function 
\begin{equation}
\tilde{F}\left(q,z\right)=\mathcal{LT}\left[F_{q}\left(t\right)\right]=\left\langle A_{-\mathbf{q}}\frac{1}{z-\hat{\Omega}}A_{\mathbf{q}}\right\rangle 
\end{equation}
where $\mathcal{LT}$ stands for Laplace transformation and $A_{\mathbf{q}}=\rho_{\mathbf{q}}/\sqrt{N}$.
One can define a projection operator on the density 
\begin{equation}
\mathcal{P}\left(\cdots\right)=A_{\mathbf{q}}\rangle\left\langle A_{\mathbf{q}}A_{-\mathbf{q}}\right\rangle ^{-1}\left\langle A_{-\mathbf{q}}\left(\cdots\right)\right\rangle 
\end{equation}
which has the property
\[
\mathcal{P}A_{\mathbf{q}}=A_{\mathbf{q}}\mbox{ and thus }\mathcal{PP=P},\mbox{ }\mathcal{P}^{n}=\mathcal{P}.
\]
Accordingly, we can define $\mathcal{Q}=\mathcal{I}-\mathcal{P}$,
which satisfies $\mathcal{Q}A_{\mathbf{q}}=0$, $\mathcal{PQ}=0$,
and $\mathcal{Q}^{n}=\mathcal{Q}$. Then for the operator $\left[z-\Omega\right]^{-1}$,
one has the following identity
\begin{equation}
\frac{1}{z-\hat{\Omega}}=\frac{1}{z-\hat{\mbox{\ensuremath{\Omega}}}\mathcal{Q}}+\frac{1}{z-\hat{\mbox{\ensuremath{\Omega}}}\mathcal{Q}}\hat{\mbox{\ensuremath{\Omega}}}\mathcal{P}\frac{1}{z-\hat{\Omega}}\label{eq:Dyson1}
\end{equation}
which is known as Dyson decomposition and can be easily checked by
right-multiplying both sides by $z-\hat{\Omega}$. 

wherein the operator $\Omega$ acts on all the functions to its right
side. In Laplace domain, this reads

\begin{eqnarray*}
\mathcal{LT}\left[\partial_{t}F\left(q,t\right)\right]\left(z\right) & = & z\tilde{F}\left(q,z\right)-F\left(q,t=0\right)=\left\langle A_{-\mathbf{q}}\hat{\Omega}\frac{1}{z-\hat{\Omega}}A_{\mathbf{q}}\right\rangle \\
 & = & \left\langle A_{-\mathbf{q}}\hat{\Omega}\mathcal{P}\frac{1}{z-\hat{\Omega}}A_{\mathbf{q}}\right\rangle +\left\langle A_{-\mathbf{q}}\hat{\Omega}\mathcal{Q}\frac{1}{z-\hat{\Omega}}A_{\mathbf{q}}\right\rangle 
\end{eqnarray*}
Using the definition of $\mathcal{P}$, the first term is 
\begin{eqnarray*}
\left\langle A_{-\mathbf{q}}\hat{\Omega}\mathcal{P}\frac{1}{z-\hat{\Omega}}A_{\mathbf{q}}\right\rangle  & = & \left\langle A_{-\mathbf{q}}\hat{\Omega}A_{\mathbf{q}}\right\rangle \left\langle A_{-\mathbf{q}}A_{\mathbf{q}}\right\rangle ^{-1}\left\langle A_{-\mathbf{q}}\frac{1}{z-\hat{\Omega}}A_{\mathbf{q}}\right\rangle \\
 & = & \left\langle A_{-\mathbf{q}}\hat{\Omega}A_{\mathbf{q}}\right\rangle \left\langle A_{-\mathbf{q}}A_{\mathbf{q}}\right\rangle ^{-1}\tilde{F}\left(q,z\right)
\end{eqnarray*}
While the second term is, using the identity(\ref{eq:Dyson1}), 
\[
\left\langle A_{-\mathbf{q}}\hat{\Omega}\mathcal{Q}\frac{1}{z-\hat{\Omega}}A_{\mathbf{q}}\right\rangle =\left\langle A_{-\mathbf{q}}\hat{\Omega}\mathcal{Q}\left[\frac{1}{z-\hat{\mbox{\ensuremath{\Omega}}}\mathcal{Q}}+\frac{1}{z-\hat{\mbox{\ensuremath{\Omega}}}\mathcal{Q}}\hat{\mbox{\ensuremath{\Omega}}}\mathcal{P}\frac{1}{z-\hat{\Omega}}\right]A_{\mathbf{q}}\right\rangle 
\]
Note that $\mathcal{Q}A_{\mathbf{q}}=0$, hence $\frac{1}{z-\hat{\mbox{\ensuremath{\Omega}}}\mathcal{Q}}A_{\mathbf{q}}=0$
and 
\begin{eqnarray*}
\left\langle A_{-\mathbf{q}}\hat{\Omega}\mathcal{Q}\frac{1}{z-\hat{\Omega}}A_{\mathbf{q}}\right\rangle  & = & \left\langle A_{-\mathbf{q}}\hat{\Omega}\mathcal{Q}\frac{1}{z-\hat{\mbox{\ensuremath{\Omega}}}\mathcal{Q}}\hat{\mbox{\ensuremath{\Omega}}}\mathcal{P}\frac{1}{z-\hat{\Omega}}A_{\mathbf{q}}\right\rangle \\
 & = & \left\langle A_{-\mathbf{q}}\hat{\Omega}\mathcal{Q}\frac{1}{z-\hat{\mbox{\ensuremath{\Omega}}}\mathcal{Q}}\hat{\mbox{\ensuremath{\Omega}}}A_{\mathbf{q}}\right\rangle \left\langle A_{-\mathbf{q}}A_{\mathbf{q}}\right\rangle ^{-1}\left\langle A_{-\mathbf{q}}\frac{1}{z-\hat{\Omega}}A_{\mathbf{q}}\right\rangle \\
 & = & \left\langle A_{-\mathbf{q}}\hat{\Omega}\mathcal{Q}\frac{1}{z-\mathcal{Q}\hat{\mbox{\ensuremath{\Omega}}}\mathcal{Q}}\mathcal{Q}\hat{\mbox{\ensuremath{\Omega}}}A_{\mathbf{q}}\right\rangle \left\langle A_{-\mathbf{q}}A_{\mathbf{q}}\right\rangle ^{-1}\tilde{F}\left(q,z\right)
\end{eqnarray*}
where we have used the definition of $\mathcal{P}$ in the second
equality and the fact $\mathcal{Q}\mathcal{Q}=\mathcal{Q}$ in the
third equality. We may introduce 
\begin{equation}
\omega_{q}=-\left\langle A_{-\mathbf{q}}\hat{\Omega}A_{\mathbf{q}}\right\rangle \left\langle A_{-\mathbf{q}}A_{\mathbf{q}}\right\rangle ^{-1}\label{eq:Omega_q-1}
\end{equation}
which is Eq.(\ref{eq:Omega_q}) and define 
\begin{equation}
\tilde{M}\left(q,z\right)=-\left\langle A_{-\mathbf{q}}\hat{\Omega}\mathcal{Q}\frac{1}{z-\mathcal{Q}\hat{\mbox{\ensuremath{\Omega}}}\mathcal{Q}}\mathcal{Q}\hat{\mbox{\ensuremath{\Omega}}}A_{\mathbf{q}}\right\rangle \left\langle A_{-\mathbf{q}}\hat{\Omega}A_{\mathbf{q}}\right\rangle ^{-1}
\end{equation}
Then the time-evolution of $F\left(q,t\right)$ in Laplace domain
reads
\[
\mathcal{LT}\left[\partial_{t}F\left(q,t\right)\right]\left(z\right)=z\tilde{F}\left(q,z\right)-F\left(q,t=0\right)=-\omega_{q}\left[1-\tilde{M}\left(q,z\right)\right]\tilde{F}\left(q,z\right)
\]
Therefore 
\begin{equation}
\tilde{F}\left(q,z\right)=\frac{F\left(q,t=0\right)}{z+\omega_{q}\left[1-\tilde{M}\left(q,z\right)\right]}
\end{equation}

For colloidal systems, there is a so-called irreducible issue \cite{2016_PRE_Szamel_AthermalActi,1995_Kawasaki_irreducible}
. Following the procedure in \cite{1996_PhysRep_Nagele_charged_suspensions}
 one needs to introduce an irreducible memory function $\tilde{M}^{{\rm irr}}\left(q,z\right)$,
which is related to$\tilde{M}\left(q,z\right)$ according to 
\begin{equation}
\tilde{M}\left(q,z\right)=\tilde{M}^{{\rm irr}}\left(q,z\right)\left[1+\tilde{M}^{{\rm irr}}\left(q,z\right)\right]^{-1}
\end{equation}
and 
\begin{equation}
\tilde{M}^{{\rm irr}}\left(q,z\right)=-\left\langle A_{-\mathbf{q}}\hat{\Omega}\mathcal{Q}\frac{1}{z-\mathcal{Q}\hat{\mbox{\ensuremath{\Omega}}}^{{\rm irr}}\mathcal{Q}}\mathcal{Q}\hat{\mbox{\ensuremath{\Omega}}}A_{\mathbf{q}}\right\rangle \left\langle A_{-\mathbf{q}}\hat{\Omega}A_{\mathbf{q}}\right\rangle ^{-1}
\end{equation}
Herein, $\hat{\Omega}^{\text{irr}}$ denotes an irreducible Smoluchowski
operator of which the detailed form is not relevant within the MCT
approximation below. Consequently, this leads to
\begin{equation}
\tilde{F}\left(q,z\right)=\frac{F\left(q,t=0\right)}{z+\frac{\omega_{q}}{1+\tilde{M}^{{\rm irr}}\left(q,z\right)}}\label{eq:GLE_ld}
\end{equation}
corresponding to in the time domain 
\begin{equation}
\frac{\partial}{\partial t}F_{q}(t)+\omega_{q}F_{q}(t)+\int_{0}^{t}duM^{{\rm irr}}\left(q,t-u\right)\frac{\partial}{\partial u}F_{q}(u)=0,\label{eq:mct}
\end{equation}
which is exactly Eq.(\ref{eq:GLE_Fq}) in the main text. Note that
above derivations are quite general and do not depend on the explicit
form of the operator $\hat{\Omega}$, whereas the expressions for
$\omega_{q}$ and $M^{\text{irr}}$ should certainly depend on $\hat{\Omega}$. 

\subsection{Frequency $\omega_{q}$}

We now substitute $A_{{\bf q}}=\rho_{\mathbf{q}}/\sqrt{N}=\sum_{j}e^{-i{\bf q}\cdot{\bf r}_{j}}/\sqrt{N}$
to calculate $\omega_{q}$. Note that 
\begin{align*}
\left\langle A_{-\mathbf{q}}\hat{\Omega}A_{\mathbf{q}}\right\rangle  & =\int d\mathbf{r}^{N}A_{-\mathbf{q}}\sum_{j}\nabla_{j}\cdot D_{j}\left[\nabla_{j}-\beta\mathbf{F}_{j}^{\text{eff}}\right]A_{\mathbf{q}}P_{s}\left(\mathbf{r}^{N}\right)\\
 & =\int d\mathbf{r}^{N}A_{-\mathbf{q}}\sum_{j}\nabla_{j}\cdot D_{j}\left\{ \nabla_{j}\left[A_{\mathbf{q}}P_{s}\left(\mathbf{r}^{N}\right)\right]-\beta\mathbf{F}_{j}^{\text{eff}}A_{\mathbf{q}}P_{s}\left(\mathbf{r}^{N}\right)\right\} 
\end{align*}
where as mentioned before, the operator $\hat{\Omega}$ acts on all
the functions to its right side including the steady-state distribution
function $P_{s}\left(\mathbf{r}^{N}\right)$. In the steady state,
the summation of all the currents $\mathbf{J}_{j}^{s}$, given by
Eq.(\ref{eq:J_i_s}), is zero according to $\hat{\Omega}P_{s}\left(\mathbf{r}^{N}\right)=-\sum_{j}\mathbf{J}_{j}^{s}=0$.
To proceed and as many authors have done, we assume more strongly
that $\mathbf{J}_{j}^{s}=0$, i.e., 
\[
\mathbf{J}_{j}^{s}=-D_{j}\left(\mathbf{r}^{N}\right)\left[\nabla_{j}-\beta\mathbf{F}_{j}^{{\rm eff}}\left(\mathbf{r}^{N}\right)\right]P_{s}\left(\mathbf{r}^{N}\right)=0.
\]
Therefore, one can obtain that 
\[
\nabla_{j}P_{s}\left(\mathbf{r}^{N}\right)=\beta\mathbf{F}_{j}^{{\rm eff}}\left(\mathbf{r}^{N}\right)P_{s}\left(\mathbf{r}^{N}\right)
\]
which is the counterpart of Yvon theorem\cite{book_Hansen_TSL} in
this nonequilibrium system.  Using this result, one has 
\begin{align*}
\left\langle A_{-\mathbf{q}}\hat{\Omega}A_{\mathbf{q}}\right\rangle  & =\int d\mathbf{r}^{N}A_{-\mathbf{q}}\sum_{j}\nabla_{j}\cdot\left[D_{j}\left(\nabla_{j}A_{\mathbf{q}}\right)P_{s}\left(\mathbf{r}^{N}\right)\right]\\
 & =-\sum_{j}\int d\mathbf{r}^{N}\left[\left(\nabla_{j}A_{-q}\right)\cdot\left(\nabla_{j}A_{\mathbf{q}}\right)\right]D_{j}P_{s}\left(\mathbf{r}^{N}\right)\\
 & =-N^{-1}\sum_{j}\int d\mathbf{r}^{N}q^{2}D_{j}P_{s}\left(\mathbf{r}^{N}\right)\\
 & =-q^{2}\sum_{j}\left\langle D_{j}\right\rangle /N=-q^{2}\bar{D}
\end{align*}
where the second equality results from partial integration and we
have used $\nabla_{j}A_{\mathbf{q}}=-i\mathbf{q}\exp\left(-i\mathbf{q}\cdot\mathbf{r}_{j}\right)/\sqrt{N}$
in the third one. $\left\langle D_{j}\right\rangle =\int d\mathbf{r}^{N}D_{j}\left(\mathbf{r}^{N}\right)P_{s}\left(\mathbf{r}^{N}\right)$
denotes the averaged instantaneous diffusion function of particle
$j$ and $\bar{D}=N^{-1}\sum_{j}\left\langle D_{j}\right\rangle $.
Therefore, the effective frequency $\omega_{q}$ reads 
\[
\omega_{q}=-\frac{\left\langle A_{-\mathbf{q}}\hat{\Omega}A_{\mathbf{q}}\right\rangle }{\left\langle A_{-\mathbf{q}}A_{\mathbf{q}}\right\rangle }=\frac{q^{2}\bar{D}}{S\left(q\right)}
\]
which is Eq.(\ref{eq:Omega_q}) in the main text.

\subsection{Memory Function $M^{\text{irr}}\left(q,t\right)$}

In the time domain, the irreducible memory function $M^{\text{irr}}\left(q,t\right)$
is given by 
\[
\tilde{M}^{{\rm irr}}\left(q,t\right)=-\left\langle A_{-\mathbf{q}}\hat{\Omega}\mathcal{Q}e^{\mathcal{Q}\hat{\Omega}^{{\rm irr}}\mathcal{Q}t}\mathcal{Q}\hat{\mbox{\ensuremath{\Omega}}}A_{\mathbf{q}}\right\rangle \left\langle A_{-\mathbf{q}}\hat{\Omega}A_{\mathbf{q}}\right\rangle ^{-1}
\]
Using the adjoint operator $\hat{\Omega}^{\dagger}$, the first term
is 
\begin{eqnarray*}
\left\langle A_{-\mathbf{q}}\hat{\Omega}\mathcal{Q}e^{\mathcal{Q}\hat{\Omega}^{{\rm irr}}\mathcal{Q}t}\mathcal{Q}\hat{\Omega}A_{\mathbf{q}}\right\rangle  & = & \left\langle A_{-\mathbf{q}}\hat{\Omega}\mathcal{Q}e^{\mathcal{Q}\hat{\Omega}^{{\rm irr}}\mathcal{Q}t}\mathcal{Q}\left(\hat{\Omega}^{\dagger}A_{\mathbf{q}}\right)\right\rangle \\
 & = & \left\langle \left(\hat{\Omega}^{\dagger}A_{-\mathbf{q}}\right)\mathcal{Q}e^{\mathcal{Q}\hat{\Omega}^{{\rm irr}}\mathcal{Q}t}\mathcal{Q}\left(\hat{\Omega}^{\dagger}A_{\mathbf{q}}\right)\right\rangle \\
 & = & \left\langle \left(\mathcal{Q}\hat{\Omega}^{\dagger}A_{-\mathbf{q}}\right)\mathcal{Q}e^{\mathcal{Q}\hat{\Omega}^{{\rm irr}}\mathcal{Q}t}\mathcal{Q}\left(\mathcal{Q}\hat{\Omega}^{\dagger}A_{\mathbf{q}}\right)\right\rangle \\
 & = & \left\langle R_{\mathbf{q}}^{*}\mathcal{Q}e^{\mathcal{Q}\hat{\Omega}^{{\rm irr}}\mathcal{Q}t}\mathcal{Q}R_{\mathbf{q}}\right\rangle 
\end{eqnarray*}
where $\mathcal{QQ=Q}$ is used in the third equality and we have
introduced
\begin{eqnarray*}
R_{\mathbf{q}} & = & \mathcal{Q}\left(\hat{\Omega}^{\dagger}A_{\mathbf{q}}\right)=(\hat{\Omega}^{\dagger}A_{\mathbf{q}})-\mathcal{P}(\Omega^{\dagger}A_{\mathbf{q}})\\
 & = & (\hat{\Omega}^{\dagger}A_{\mathbf{q}})-\frac{\left\langle A_{-\mathbf{q}}\left(\hat{\Omega}^{\dagger}A_{\mathbf{q}}\right)\right\rangle }{\left\langle A_{-\mathbf{q}}A_{\mathbf{q}}\right\rangle }A_{\mathbf{q}}\\
 & = & =(\hat{\Omega}^{\dagger}A_{\mathbf{q}})-\frac{\left\langle A_{-\mathbf{q}}\hat{\Omega}A_{\mathbf{q}}\right\rangle }{\left\langle A_{-\mathbf{q}}A_{\mathbf{q}}\right\rangle }A_{\mathbf{q}}\\
 & = & (\hat{\Omega}^{\dagger}A_{\mathbf{q}})+\omega_{q}A_{\mathbf{q}}
\end{eqnarray*}
which is a type of ``random force''. 

It is in this step one needs to introduce the mode-coupling approximation.
The memory function is assumed to be dominated by the projection onto
the coupling density modes \cite{book_gotze2008complex}. One can
then define a second-order projection operator
\begin{equation}
\mathcal{P}_{2}\equiv\frac{1}{2}\sum_{{\bf k},{\bf p}}\left|A_{\mathbf{p}}A_{\mathbf{k}}\right\rangle \left\langle A_{\mathbf{p}}^{*}A_{\mathbf{k}}^{*}A_{\mathbf{p}}A_{\mathbf{k}}\right\rangle ^{-1}\left\langle A_{\mathbf{p}}A_{\mathbf{k}}\right|
\end{equation}
and make the approximation:
\begin{eqnarray}
\left\langle R_{\mathbf{q}}^{*}e^{\mathcal{Q}\hat{\Omega}^{{\rm irr}}\mathcal{Q}t}R_{\mathbf{q}}\right\rangle  & \approx & \left\langle R_{\mathbf{q}}^{*}\mathcal{P}_{2}e^{\mathcal{Q}\hat{\Omega}^{{\rm irr}}\mathcal{Q}t}\mathcal{P}_{2}R_{\mathbf{q}}\right\rangle \nonumber \\
 & = & \frac{1}{4}\sum_{{\bf k},{\bf p}}\sum_{{\bf k}',{\bf p}'}\left\langle R_{\mathbf{q}}^{*}A_{\mathbf{p}}A_{\mathbf{k}}\right\rangle \left\langle A_{\mathbf{p}}^{*}A_{\mathbf{k}}^{*}A_{\mathbf{p}}A_{\mathbf{k}}\right\rangle ^{-1}\nonumber \\
 &  & \times\left\langle A_{\mathbf{p}'}^{*}A_{\mathbf{k}'}^{*}R_{\mathbf{q}}\right\rangle \left\langle A_{\mathbf{p}'}^{*}A_{\mathbf{k}'}^{*}A_{\mathbf{p}'}A_{\mathbf{k}'}\right\rangle ^{-1}\nonumber \\
 &  & \times\left\langle A_{\mathbf{p}}A_{\mathbf{k}}e^{\mathcal{Q}\hat{\Omega}^{{\rm irr}}\mathcal{Q}t}A_{\mathbf{p}'}A_{\mathbf{k}'}\right\rangle \nonumber \\
 & \approx & \frac{1}{4}\sum_{{\bf k},{\bf p}}\sum_{{\bf k}',{\bf p}'}\frac{\left\langle \rho_{\mathbf{p}'}^{*}\rho_{\mathbf{k}'}^{*}R_{\mathbf{q}}\right\rangle }{NS\left(k'\right)S\left(p'\right)}\frac{\left\langle R_{\mathbf{q}}^{*}\rho_{\mathbf{p}}\rho_{\mathbf{k}}\right\rangle }{NS\left(k\right)S\left(p\right)}\nonumber \\
 &  & \times\frac{1}{N^{2}}\left[\delta_{{\bf pp}'}\delta_{{\bf kk}'}\left\langle \rho_{\mathbf{p}}^{*}e^{\hat{\Omega}t}\rho_{\mathbf{p}'}\right\rangle \left\langle \rho_{\mathbf{k}}^{*}e^{\hat{\Omega}t}\rho_{\mathbf{k}'}\right\rangle \right.\nonumber \\
 &  & +\left.\delta_{{\bf pk}'}\delta_{{\bf kp}'}\left\langle \rho_{\mathbf{p}}^{*}e^{\hat{\Omega}t}\rho_{\mathbf{k}'}\right\rangle \left\langle \rho_{\mathbf{k}}^{*}e^{\hat{\Omega}t}\rho_{\mathbf{p}'}\right\rangle \right]\nonumber \\
 & = & \frac{1}{2}\sum_{{\bf k},{\bf p}}\frac{\left|\left\langle \rho_{\mathbf{p}}^{*}\rho_{\mathbf{k}}^{*}R_{\mathbf{q}}\right\rangle \right|^{2}}{\left[N^{2}S\left(k\right)S\left(p\right)\right]^{2}}\left\langle \rho_{\mathbf{p}}^{*}e^{\hat{\Omega}t}\rho_{\mathbf{p}}\right\rangle \left\langle \rho_{\mathbf{k}}^{*}e^{\hat{\Omega}t}\rho_{\mathbf{k}}\right\rangle \label{eq:RqRq_t}
\end{eqnarray}
Here we have to factories the static and dynamic four-point correlation
functions into products of two-point functions
\begin{eqnarray*}
\left\langle \rho_{\mathbf{p}}^{*}\rho_{\mathbf{k}}^{*}\rho_{\mathbf{p}'}\rho_{\mathbf{k}'}\right\rangle  & \approx & \left\langle \rho_{\mathbf{p}}^{*}\rho_{\mathbf{k}'}\right\rangle \left\langle \rho_{\mathbf{k}}^{*}\rho_{\mathbf{p}'}\right\rangle +\left\langle \rho_{\mathbf{p}}^{*}\rho_{\mathbf{p}'}\right\rangle \left\langle \rho_{\mathbf{k}}^{*}\rho_{\mathbf{k}'}\right\rangle \\
 & = & \delta_{{\bf p},{\bf k}'}\delta_{{\bf k},{\bf p}'}N^{2}S(p)S(k)+\delta_{{\bf p},{\bf p}'}\delta_{{\bf k},{\bf k}'}N^{2}S(p)S(k)
\end{eqnarray*}
and simultaneously replace the projected operator $\mathcal{Q}\hat{\Omega}^{{\rm irr}}\mathcal{Q}$
by the full Smoluchowski operator $\hat{\Omega}$ in the propagator
governing the time evolution of the correlation function \cite{book_Hansen_TSL}
\[
\left\langle A_{\mathbf{p}}A_{\mathbf{k}}e^{\mathcal{Q}\hat{\Omega}^{{\rm irr}}\mathcal{Q}t}A_{\mathbf{p}'}A_{\mathbf{k}'}\right\rangle \approx\delta_{{\bf pp}'}\delta_{{\bf kk}'}\left\langle \rho_{\mathbf{p}}^{*}e^{\hat{\Omega}t}\rho_{\mathbf{p}'}\right\rangle \left\langle \rho_{\mathbf{k}}^{*}e^{\hat{\Omega}t}\rho_{\mathbf{k}'}\right\rangle +\delta_{{\bf pk}'}\delta_{{\bf kp}'}\left\langle \rho_{\mathbf{p}}^{*}e^{\hat{\Omega}t}\rho_{\mathbf{k}'}\right\rangle \left\langle \rho_{\mathbf{k}}^{*}e^{\hat{\Omega}t}\rho_{\mathbf{p}'}\right\rangle 
\]

We now need to calculate $\left\langle \rho_{\mathbf{p}}^{*}\rho_{\mathbf{k}}^{*}R_{\mathbf{q}}\right\rangle $,
which is given by 
\begin{equation}
\left\langle \rho_{\mathbf{p}}^{*}\rho_{\mathbf{k}}^{*}R_{\mathbf{q}}\right\rangle =\frac{1}{\sqrt{N}}\left[\left\langle \left(\rho_{\mathbf{p}}\rho_{\mathbf{k}}\right)^{*}(\hat{\Omega}^{\dagger}\rho_{\mathbf{q}})\right\rangle +\omega_{q}\left\langle \left(\rho_{\mathbf{p}}\rho_{\mathbf{k}}\right)^{*}\rho_{\mathbf{q}}\right\rangle \right]\label{eq:rhorhoRq}
\end{equation}
The first term in the bracket is 
\begin{eqnarray*}
\left\langle \left(\rho_{\mathbf{p}}\rho_{\mathbf{k}}\right)^{*}(\hat{\Omega}^{\dagger}\rho_{\mathbf{q}})\right\rangle  & = & \left\langle \left(\rho_{\mathbf{p}}\rho_{\mathbf{k}}\right)^{*}\hat{\Omega}\rho_{\mathbf{q}}\right\rangle \\
 & = & \sum_{j}\left\langle \left(-i\mathbf{p}e^{i\mathbf{p}\cdot\mathbf{r}_{j}}\rho_{\mathbf{k}}-i\mathbf{k}e^{i\mathbf{k}\cdot\mathbf{r}_{j}}\rho_{\mathbf{p}}\right)\cdot D_{j}\left(-i\mathbf{q}e^{-i\mathbf{q}\cdot\mathbf{r}_{j}}\right)\right\rangle \\
 & = & -\mathbf{q}\cdot\mathbf{p}\left\langle \sum_{j,l}D_{j}e^{-i\left(\mathbf{q}-\mathbf{p}\right)\cdot\mathbf{r}_{j}+i\mathbf{k}\cdot\mathbf{r}_{l}}\right\rangle -\mathbf{q}\cdot\mathbf{k}\left\langle \sum_{j,l}D_{j}e^{-i\left(\mathbf{q}-\mathbf{k}\right)\cdot\mathbf{r}_{j}+i\mathbf{p}\cdot\mathbf{r}_{l}}\right\rangle \\
 & = & -\mathbf{q}\cdot\mathbf{p}\delta_{\mathbf{q},\mathbf{k}+\mathbf{p}}\left[\left\langle \sum_{j,l}D_{j}e^{-i\mathbf{k}\cdot\mathbf{r}_{j}+i\mathbf{k}\cdot\mathbf{r}_{l}}\right\rangle +\left\langle \sum_{j,l}D_{j}e^{-i\mathbf{p}\cdot\mathbf{r}_{j}+i\mathbf{p}\cdot\mathbf{r}_{l}}\right\rangle \right]\\
 & = & -ND_{0}\delta_{\mathbf{q},\mathbf{k}+\mathbf{p}}\left[\left(\mathbf{q}\cdot\mathbf{p}\right)S_{2}(k)+\left(\mathbf{q}\cdot\mathbf{k}\right)S_{2}\left(p\right)\right]
\end{eqnarray*}
where the second equality is simply a result of partial integration,
and the fourth equality results from translational invariance. For
short of notation, we have introduced a function $S_{2}\left(k\right)$
in the fourth equality defined as 
\[
S_{2}\left(k\right)=\frac{1}{ND_{0}}\left\langle \sum_{j,l}D_{j}e^{-i\mathbf{k}\cdot\left(\mathbf{r}_{j}-\mathbf{r}_{l}\right)}\right\rangle 
\]
in accordance with Eq.(\ref{eq:S2_k}) in the main text. If $D_{j}$
is a constant, such as$D_{j}=D_{0}$ in the $\tau_{p}\rightarrow0$
limit, it can be drawn out of the bracket such that $S_{2}\left(k\right)=N^{-1}\left\langle \sum_{j,l}e^{-i\mathbf{k}\cdot\left(\mathbf{r}_{j}-\mathbf{r}_{l}\right)}\right\rangle =S\left(k\right)$
which is exactly the static structure factor. Nevertheless, in our
present case, $D_{j}$ depends on the instantaneous configuration
$\mathbf{r}^{N}$, such that $S_{2}\left(k\right)$ may show abundant
features different from $S\left(k\right)$. It is interesting to note
that for large $k$, $D_{j}$ seems to be decoupled from the Fourier
components $\exp\left(-i\mathbf{k}\cdot\mathbf{r}_{l}\right)$, and
$S_{2}\left(k\right)$ can be approximated by 
\begin{equation}
S_{2}\left(k\right)\simeq\frac{\bar{D}}{ND_{0}}\left\langle \sum_{j,l}e^{-i\mathbf{k}\cdot\left(\mathbf{r}_{j}-\mathbf{r}_{l}\right)}\right\rangle =\frac{\bar{D}}{D_{0}}S\left(k\right)\label{eq:s2approxDsk}
\end{equation}
as shown in Fig.\ref{Fig_S2k} in the main text.

The second term in Eq.(\ref{eq:rhorhoRq}) can be calculated using
the so-called convolution approximation, which is assumed to be still
appropriate in nonequilibrium situation\cite{2014_PRE_Szamel_EffecTemp},
\begin{equation}
\left\langle \rho_{\mathbf{p}}^{*}\rho_{\mathbf{k}}^{*}\rho_{\mathbf{q}}\right\rangle \approx\delta_{\mathbf{k}+\mathbf{p},\mathbf{q}}NS(q)S(p)S(k)
\end{equation}
Therefore, we can get
\begin{eqnarray*}
\left\langle \rho_{\mathbf{p}}^{*}\rho_{\mathbf{k}}^{*}R_{\mathbf{q}}\right\rangle  & = & -\sqrt{N}D_{0}\delta_{\mathbf{q},\mathbf{k}+\mathbf{p}}\left[\mathbf{q}\cdot\mathbf{p}S_{2}(k)+\mathbf{q}\cdot\mathbf{k}S_{2}\left(p\right)-q^{2}\frac{\bar{D}}{D_{0}}S(p)S(k)\right]
\end{eqnarray*}
Substituting this into Eq.(\ref{eq:RqRq_t}), we obtain
\begin{align*}
\left\langle R_{\mathbf{q}}^{*}e^{\mathcal{Q}\hat{\Omega}^{{\rm irr}}\mathcal{Q}t}R_{\mathbf{q}}\right\rangle  & \approx\frac{1}{2}\sum_{{\bf k},{\bf p}}\frac{\left|\left\langle \rho_{\mathbf{p}}^{*}\rho_{\mathbf{k}}^{*}R_{\mathbf{q}}\right\rangle \right|^{2}}{\left[N^{2}S\left(k\right)S\left(p\right)\right]^{2}}\left\langle \rho_{\mathbf{p}}^{*}e^{\hat{\Omega}t}\rho_{\mathbf{p}}\right\rangle \left\langle \rho_{\mathbf{k}}^{*}e^{\hat{\Omega}t}\rho_{\mathbf{k}}\right\rangle \\
 & =\frac{1}{2N}\sum_{{\bf k},{\bf p}}\left|V_{\mathbf{q}}\left(\mathbf{k},\mathbf{p}\right)\right|^{2}\left\langle \rho_{\mathbf{p}}^{*}e^{\hat{\Omega}t}\rho_{\mathbf{p}}\right\rangle \left\langle \rho_{\mathbf{k}}^{*}e^{\hat{\Omega}t}\rho_{\mathbf{k}}\right\rangle 
\end{align*}
where the vortex function $V_{\mathbf{q}}\left(\mathbf{k},\mathbf{p}\right)$
is defined as

\begin{eqnarray*}
V_{\mathbf{q}}\left(\mathbf{k},\mathbf{p}\right) & = & \sqrt{N}\left\langle \rho_{\mathbf{p}}^{*}\rho_{\mathbf{k}}^{*}R_{\mathbf{q}}\right\rangle \left[N^{2}S\left(k\right)S\left(p\right)\right]^{-1}\\
 & = & -\delta_{\mathbf{k}+\mathbf{p},\mathbf{q}}\frac{D_{0}}{N}\left\{ \left(\mathbf{q}\cdot\mathbf{k}\right)\frac{S_{2}(p)}{S(p)S(k)}+\left(\mathbf{q}\cdot\mathbf{p}\right)\frac{S_{2}(k)}{S(p)S(k)}-q^{2}\frac{\bar{D}}{D_{0}}\right\} \\
 & = & -\delta_{\mathbf{k}+\mathbf{p},\mathbf{q}}\frac{\bar{D}}{N}\left\{ \left(\mathbf{q}\cdot\mathbf{k}\right)\left[\frac{D_{0}S_{2}(p)}{\bar{D}S(p)S(k)}-1\right]+\left(\mathbf{q}\cdot\mathbf{p}\right)\left[\frac{D_{0}S_{2}(k)}{\bar{D}S(p)S(k)}-1\right]\right\} 
\end{eqnarray*}
Now it is instructive for us to define a pseudo ``direct correlation
function'' as 
\[
C_{2}(\mathbf{q};\mathbf{k})=\frac{\delta_{\mathbf{k}+\mathbf{p},\mathbf{q}}}{\rho}\left[1-\frac{D_{0}S_{2}(p)}{\bar{D}S(p)S(k)}\right]\equiv\frac{1}{\rho}\left[1-\frac{D_{0}S_{2}(\left|\mathbf{q}-\mathbf{k}\right|)}{\bar{D}S(\left|\mathbf{q}-\mathbf{k}\right|)S(k)}\right]
\]
Note that if Eq.(\ref{eq:s2approxDsk}) becomes a equality, namely,
$S_{2}(k)=\bar{D}S(k)/D_{0}$, then
\[
C_{2}({\bf q};{\bf k})=\frac{1}{\rho}\left[1-\frac{1}{S(k)}\right]=c(k)
\]
which has the same form as the conventional direct correlation function.
With this notation, the vortex can be written as
\[
V_{\mathbf{q}}\left(\mathbf{k},\mathbf{p}\right)=\frac{\rho\bar{D}}{N}\left[\left(\mathbf{q}\cdot\mathbf{k}\right)C_{2}\left(\mathbf{q};\mathbf{k}\right)+\left(\mathbf{q}\cdot\mathbf{p}\right)C_{2}(\mathbf{q};\mathbf{k})\right]
\]
and 
\begin{align}
\left\langle R_{\mathbf{q}}^{*}e^{\mathcal{Q}\hat{\Omega}^{{\rm irr}}\mathcal{Q}t}R_{\mathbf{q}}\right\rangle  & =\frac{1}{2N}\sum_{{\bf k},{\bf p}}\left|V_{\mathbf{q}}\left(\mathbf{k},\mathbf{p}\right)\right|^{2}\left\langle \rho_{\mathbf{p}}^{*}e^{\hat{\Omega}t}\rho_{\mathbf{p}}\right\rangle \left\langle \rho_{\mathbf{k}}^{*}e^{\hat{\Omega}t}\rho_{\mathbf{k}}\right\rangle \nonumber \\
 & \approx\frac{1}{2}\sum_{\mathbf{k}}\frac{\rho^{2}\bar{D}^{2}}{N}\left[\left(\mathbf{q}\cdot\mathbf{k}\right)C_{2}\left({\bf q};{\bf k}\right)+\left(\mathbf{q}\cdot\mathbf{p}\right)C_{2}({\bf q};{\bf p})\right]^{2}F_{k}(t)F_{p}(t)
\end{align}
Consequently, we get the irreducible memory function 
\begin{eqnarray*}
\tilde{M}^{{\rm irr}}\left(q;t\right) & \approx & -\left\langle R_{\mathbf{q}}^{*}e^{\mathcal{Q}\hat{\Omega}^{{\rm irr}}\mathcal{Q}t}R_{\mathbf{q}}\right\rangle \left\langle A_{-\mathbf{q}}\hat{\Omega}A_{\mathbf{q}}\right\rangle ^{-1}\\
 & = & \frac{\rho^{2}\bar{D}}{2q^{2}N}\sum_{\mathbf{k}}\left[\left(\mathbf{q}\cdot\mathbf{k}\right)C_{2}\left({\bf q};{\bf k}\right)+\left(\mathbf{q}\cdot\mathbf{p}\right)C_{2}({\bf q};{\bf p})\right]^{2}F_{k}(t)F_{p}(t)
\end{eqnarray*}
Changing to integration by using $\sum_{\mathbf{k}}\to\left(2\pi\right)^{-3}V\int d^{3}\mathbf{k}$,
one has
\begin{equation}
\tilde{M}^{{\rm irr}}\left(q;t\right)=\frac{\rho\bar{D}}{16\pi^{3}}\int d^{3}\mathbf{k}\left[\left(\hat{\mathbf{q}}\cdot\mathbf{k}\right)C_{2}\left({\bf q};{\bf k}\right)+\left(\hat{\mathbf{q}}\cdot\mathbf{p}\right)C_{2}({\bf q};{\bf p})\right]^{2}F_{k}(t)F_{p}(t)
\end{equation}
where $\mathbf{\hat{q}}=\mathbf{q}/q$ is the unit vector in the direction
of $\mathbf{q}$. This is Eq. (\ref{eq:Memory_Irr}) in the main text.

\subsection{Tagged Particle Dynamics}

We now consider tagged particle dynamics. The relevant variable is
$A_{\mathbf{q}}=\rho_{\mathbf{q}}^{s}=e^{-i\mathbf{q}\cdot\mathbf{r}_{s}}$
( the subscript or superscript 's' stands for the single particle)
and the self-scattering function reads $F_{q}^{s}\left(t\right)=\left\langle \rho_{-\mathbf{q}}^{s}e^{\hat{\Omega}t}\rho_{\mathbf{q}}^{s}\right\rangle $
with $F_{q}^{s}\left(0\right)=1$. The derivation of the memory function
equation for $F_{q}^{s}\left(t\right)$ are similar to that of $F_{q}\left(t\right)$,
except that some relevant calculations are different. Briefly, we
have 
\[
M_{s}^{\text{irr}}\left(q,t\right)=-\left\langle R_{\mathbf{q}}^{s*}e^{\mathcal{Q}\hat{\Omega}^{{\rm irr}}\mathcal{Q}t}R_{\mathbf{q}}^{s}\right\rangle \left\langle \rho_{-\mathbf{q}}^{s}\hat{\Omega}\rho_{\mathbf{q}}^{s}\right\rangle ^{-1}
\]
where 
\begin{eqnarray*}
R_{\mathbf{q}}^{s} & = & \mathcal{Q}\left(\Omega^{\dagger}\rho_{\mathbf{q}}^{s}\right)=(\Omega^{\dagger}\rho_{\mathbf{q}}^{s})-\mathcal{P}(\Omega^{\dagger}\rho_{\mathbf{q}}^{s})\\
 & = & (\Omega^{\dagger}\rho_{\mathbf{q}}^{s})-\frac{\left\langle \rho_{-\mathbf{q}}^{s}\left(\hat{\Omega}^{\dagger}\rho_{\mathbf{q}}^{s}\right)\right\rangle }{\left\langle \rho_{-\mathbf{q}}^{s}\rho_{\mathbf{q}}^{s}\right\rangle }\rho_{\mathbf{q}}^{s}\\
 & = & (\Omega^{\dagger}\rho_{\mathbf{q}}^{s})+q^{2}\bar{D}\rho_{\mathbf{q}}^{s}=(\Omega^{\dagger}\rho_{\mathbf{q}}^{s})
\end{eqnarray*}
In the third equality of above equation for $R_{\mathbf{q}}^{s}$,
we have used the result $\left\langle \rho_{-\mathbf{q}}^{s}\hat{\Omega}\rho_{\mathbf{q}}^{s}\right\rangle =-q^{2}\left\langle D_{s}\right\rangle $. 

To calculate $\left\langle R_{\mathbf{q}}^{s*}e^{\mathcal{Q}\hat{\Omega}^{{\rm irr}}\mathcal{Q}t}R_{\mathbf{q}}^{s}\right\rangle $,
one projects $R_{\mathbf{q}}^{s}$ onto the product of single and
collective modes $\rho_{\mathbf{k}}\rho_{\mathbf{p}}^{s}$ with projection
operator
\begin{equation}
\mathcal{P}_{2}^{s}=\sum_{\mathbf{k},\mathbf{p}}\left.\rho_{\mathbf{k}}\rho_{\mathbf{p}}^{s}\right\rangle \left\langle \left(\rho_{\mathbf{k}}\rho_{\mathbf{p}}^{s}\right)^{*}\left(\rho_{\mathbf{k}}\rho_{\mathbf{p}}^{s}\right)\right\rangle ^{-1}\left\langle \left(\rho_{\mathbf{k}}\rho_{\mathbf{p}}^{s}\right)^{*}\right.
\end{equation}
Then
\begin{eqnarray}
\left\langle R_{-\mathbf{q}}^{s}e^{\mathcal{Q}\hat{\Omega}^{{\rm irr}}\mathcal{Q}t}R_{\mathbf{q}}^{s}\right\rangle  & \approx & \left\langle R_{-\mathbf{q}}^{s}\mathcal{P}_{2}^{s}e^{\mathcal{Q}\hat{\Omega}^{{\rm irr}}\mathcal{Q}t}\mathcal{P}_{2}^{s}R_{\mathbf{q}}^{s}\right\rangle \nonumber \\
 & = & \sum_{\mathbf{k},\mathbf{p}}\sum_{\mathbf{k}',\mathbf{p}'}\left\langle R_{-\mathbf{q}}^{s}\left(\rho_{\mathbf{k}}\rho_{\mathbf{p}}^{s}\right)\right\rangle \left\langle \left(\rho_{\mathbf{k}}\rho_{\mathbf{p}}^{s}\right)^{*}\left(\rho_{\mathbf{k}}\rho_{\mathbf{p}}^{s}\right)\right\rangle ^{-1}\nonumber \\
 &  & \times\left\langle \left(\rho_{\mathbf{k}'}\rho_{\mathbf{p}'}^{s}\right)^{*}R_{\mathbf{q}}^{s}\right\rangle \left\langle \left(\rho_{\mathbf{k}'}\rho_{\mathbf{p}'}^{s}\right)^{*}\left(\rho_{\mathbf{k}'}\rho_{\mathbf{p}'}^{s}\right)\right\rangle ^{-1}\left\langle \left(\rho_{\mathbf{k}}\rho_{\mathbf{p}}^{s}\right)^{*}e^{\mathcal{Q}\hat{\Omega}^{{\rm irr}}\mathcal{Q}t}\left(\rho_{\mathbf{k}'}\rho_{\mathbf{p}'}^{s}\right)\right\rangle \nonumber \\
 & \simeq & \sum_{\mathbf{k},\mathbf{p}}\sum_{\mathbf{k}',\mathbf{p}'}\frac{\left\langle R_{-\mathbf{q}}^{s}\left(\rho_{\mathbf{k}}\rho_{\mathbf{p}}^{s}\right)\right\rangle }{NS\left(k\right)}\frac{\left\langle \left(\rho_{\mathbf{k}'}\rho_{\mathbf{p}'}^{s}\right)^{*}R_{\mathbf{q}}^{s}\right\rangle }{NS\left(k'\right)}\delta_{\mathbf{kk}'}\delta_{\mathbf{pp}'}\left\langle \rho_{\mathbf{k}}^{*}e^{\hat{\Omega}t}\rho_{\mathbf{k}'}\right\rangle \left\langle \rho_{\mathbf{p}}^{s*}e^{\hat{\Omega}t}\rho_{\mathbf{p}'}^{s}\right\rangle \nonumber \\
 & = & \sum_{\mathbf{k},\mathbf{p}}\left|V_{q}^{s}\left(\mathbf{k},\mathbf{p}\right)\right|^{2}\left\langle \rho_{-\mathbf{k}}e^{\hat{\Omega}t}\rho_{\mathbf{k}}\right\rangle \left\langle \rho_{-\mathbf{p}}^{s}e^{\hat{\Omega}t}\rho_{\mathbf{p}}^{s}\right\rangle 
\end{eqnarray}
where

\[
V_{\mathbf{q}}^{s}\left(\mathbf{k},\mathbf{p}\right)=\frac{\left\langle \left(\rho_{\mathbf{k}}\rho_{\mathbf{p}}^{s}\right)^{*}R_{\mathbf{q}}^{s}\right\rangle }{NS\left(k\right)}
\]
Now we need to calculate $\left\langle \left(\rho_{\mathbf{k}}\rho_{\mathbf{p}}^{s}\right)^{*}R_{\mathbf{q}}^{s}\right\rangle $,
which is given by 
\[
\left\langle \left(\rho_{\mathbf{k}}\rho_{\mathbf{p}}^{s}\right)^{*}R_{\mathbf{q}}^{s}\right\rangle =\left\langle \left(\rho_{\mathbf{k}}\rho_{\mathbf{p}}^{s}\right)^{*}(\Omega^{\dagger}\rho_{\mathbf{q}}^{s})\right\rangle +q^{2}\bar{D}\left\langle \left(\rho_{\mathbf{k}}\rho_{\mathbf{p}}^{s}\right)^{*}\rho_{\mathbf{q}}^{s}\right\rangle 
\]

The second term is 
\begin{align*}
\left\langle \left(\rho_{\mathbf{k}}\rho_{\mathbf{p}}^{s}\right)^{*}\rho_{\mathbf{q}}^{s}\right\rangle  & =\left\langle \rho_{-\mathbf{k}}\rho_{\mathbf{q}-\mathbf{p}}^{s}\right\rangle =\delta_{\mathbf{q},\mathbf{k}+\mathbf{p}}\rho c\left(k\right)S\left(k\right)=\delta_{\mathbf{q},\mathbf{k}+\mathbf{p}}\left[S\left(k\right)-1\right]
\end{align*}
where the $\delta$ symbol results from translational invariance and
the result $\left\langle \rho_{\mathbf{k}}^{*}\rho_{\mathbf{k}}^{s}\right\rangle =\rho c\left(k\right)S\left(k\right)=S\left(k\right)-1$\cite{book_gotze2008complex}.

The first term is
\begin{eqnarray*}
\left\langle \left(\rho_{\mathbf{k}}\rho_{\mathbf{p}}^{s}\right)^{*}(\Omega^{\dagger}\rho_{\mathbf{q}}^{s})\right\rangle  & = & \left\langle \left(\rho_{\mathbf{k}}\rho_{\mathbf{p}}^{s}\right)^{*}\hat{\Omega}\rho_{\mathbf{q}}^{s}\right\rangle \\
 & = & \sum_{j=1}^{N}\left\langle \left(-i\mathbf{k}\left(\sum_{l\neq s}\delta_{jl}e^{i\mathbf{k}\cdot\mathbf{r}_{l}}\right)e^{i\mathbf{p}\cdot\mathbf{r}_{s}}-\delta_{js}i\mathbf{p}\rho_{-\mathbf{k}}e^{i\mathbf{p}\cdot\mathbf{r}_{s}}\right)D_{j}\cdot\left(-\delta_{js}i\mathbf{q}e^{-i\mathbf{q}\cdot\mathbf{r}_{s}}\right)\right\rangle \\
 & = & -\left\langle \left(\mathbf{p}\cdot\mathbf{q}\right)D_{s}\rho_{-\mathbf{k}}e^{i\left(\mathbf{p}-\mathbf{q}\right)\cdot\mathbf{r}_{s}}\right\rangle \\
 & = & -\delta_{\mathbf{q},\mathbf{k}+\mathbf{p}}\left(\mathbf{p}\cdot\mathbf{q}\right)\left(D_{0}S_{2}\left(k\right)-\bar{D}\right)
\end{eqnarray*}
where we have used partial integration and Yvon theorem in the second
equality. 

\begin{eqnarray*}
\left\langle \left(\rho_{k}\rho_{p}^{s}\right)^{*}R_{\mathbf{q}}^{s}\right\rangle  & = & -\delta_{\mathbf{q},\mathbf{k}+\mathbf{p}}\left[\mathbf{p}\cdot\mathbf{q}\left(D_{0}S_{2}\left(k\right)-\bar{D}\right)\right]+q^{2}\bar{D}\delta_{\mathbf{q},\mathbf{k}+\mathbf{p}}\left(S(k)-1\right)\\
 & = & \delta_{\mathbf{q},\mathbf{k}+\mathbf{p}}\bar{D}\left[\mathbf{k}\cdot\mathbf{q}\left(S(k)-1\right)+\mathbf{p}\cdot\mathbf{q}\left(S(k)-\frac{D_{0}S_{2}\left(k\right)}{\bar{D}}\right)\right]
\end{eqnarray*}
\begin{eqnarray*}
V_{q}^{s}\left(\mathbf{k},\mathbf{p}\right) & = & \frac{\left\langle \left(\rho_{\mathbf{k}}\rho_{\mathbf{p}}^{s}\right)^{*}R_{\mathbf{q}}^{s}\right\rangle }{NS(k)}=\delta_{\mathbf{q},\mathbf{k}+\mathbf{p}}\frac{\bar{D}}{N}\left[\mathbf{k}\cdot\mathbf{q}\left(1-\frac{1}{S(k)}\right)+\mathbf{p}\cdot\mathbf{q}\left(1-\frac{D_{0}S_{2}\left(k\right)}{\bar{D}S(k)}\right)\right]\\
 & = & \delta_{\mathbf{q},\mathbf{k}+\mathbf{p}}\frac{\rho\bar{D}}{N}\left[\mathbf{k}\cdot\mathbf{q}c(k)+\mathbf{p}\cdot\mathbf{q}\frac{1}{\rho}\left(1-\frac{D_{0}S_{2}\left(k\right)}{\bar{D}S(k)}\right)\right]
\end{eqnarray*}
Notice if $S_{2}(k)=\bar{D}S(k)/D_{0}$, $V_{q}\left(\mathbf{k},\mathbf{p}\right)=\delta_{\mathbf{q},\mathbf{k}+\mathbf{p}}\frac{\rho\bar{D}}{N}\left[\mathbf{k}\cdot\mathbf{q}c(k)\right]$,
which is the equilibrium result. Next
\begin{align}
\left\langle R_{\mathbf{q}}^{s*}e^{\mathcal{Q}\hat{\Omega}^{{\rm irr}}\mathcal{Q}t}R_{\mathbf{q}}^{s}\right\rangle  & =\sum_{\mathbf{k},\mathbf{p}}\left|V_{q}^{s}\left(\mathbf{k},\mathbf{p}\right)\right|^{2}\left\langle \rho_{-\mathbf{k}}e^{\hat{\Omega}t}\rho_{\mathbf{k}}\right\rangle \left\langle \rho_{-\mathbf{p}}^{s}e^{\hat{\Omega}t}\rho_{\mathbf{p}}^{s}\right\rangle \\
 & \approx\sum_{\mathbf{k}}\frac{\rho^{2}\bar{D}^{2}}{N}\left[\mathbf{k}\cdot\mathbf{q}c(k)+\mathbf{p}\cdot\mathbf{q}\frac{1}{\rho}\left(1-\frac{D_{0}S_{2}\left(k\right)}{\bar{D}S(k)}\right)\right]^{2}F_{k}(t)F_{p}^{s}(t)
\end{align}
and
\begin{eqnarray}
\tilde{M}_{s}^{{\rm irr}}\left(q;t\right) & = & -\left\langle R_{\mathbf{q}}^{s*}e^{\mathcal{Q}\hat{\Omega}^{{\rm irr}}\mathcal{Q}t}R_{\mathbf{q}}^{s}\right\rangle \left\langle \rho_{-\mathbf{q}}^{s}\hat{\Omega}\rho_{\mathbf{q}}^{s}\right\rangle ^{-1}\nonumber \\
 & \approx & \frac{\rho^{2}\bar{D}}{q^{2}N}\sum_{\mathbf{k}}\left[\mathbf{k}\cdot\mathbf{q}c(k)+\mathbf{p}\cdot\mathbf{q}\frac{1}{\rho}\left(1-\frac{D_{0}S_{2}\left(k\right)}{\bar{D}S(k)}\right)\right]^{2}F_{k}(t)F_{p}^{s}(t)\\
 & = & \frac{\rho\bar{D}}{\left(2\pi\right)^{3}}\int d^{3}\mathbf{k}\left[\mathbf{k}\cdot\hat{\mathbf{q}}c(k)+\mathbf{p}\cdot\hat{\mathbf{q}}\frac{1}{\rho}\left(1-\frac{D_{0}S_{2}\left(k\right)}{\bar{D}S(k)}\right)\right]^{2}F_{k}(t)F_{p}^{s}(t)
\end{eqnarray}
Finally, the memory function equation for the self-scattering function
$F_{q}^{s}\left(t\right)$ is given by
\begin{equation}
\frac{\partial}{\partial t}F_{q}^{s}(t)+\omega_{q}^{s}F_{q}^{s}(t)+\int_{0}^{t}duM_{s}^{{\rm irr}}\left(q,t-u\right)\frac{\partial}{\partial u}F_{q}^{s}(u)=0
\end{equation}
where 
\begin{equation}
\omega_{q}^{s}=-\frac{\left\langle \rho_{-\mathbf{q}}^{s}\left(\hat{\Omega}^{\dagger}\rho_{\mathbf{q}}^{s}\right)\right\rangle }{\left\langle \rho_{-\mathbf{q}}^{s}\rho_{\mathbf{q}}^{s}\right\rangle }=q^{2}\bar{D}.
\end{equation}

\end{document}